\documentclass{article}

\usepackage[utf8]{inputenc} % Character encoding

\usepackage[frozencache=true,cachedir=minted-cache]{minted} 

\usepackage{mathrsfs}
\usepackage{caption}
\usepackage{subcaption}
\usepackage{listings}
\usepackage{appendix}
\newcommand{\addappendix}{   % Self-created command to insert appendix with predefined settings
    \newpage
    \appendix
    \section*{Appendix}   % Name of appendix
    \addcontentsline{toc}{section}{Appendix}  % Add appendix name to table of contents
    \renewcommand{\thesubsection}{\Alph{subsection}}    % Change numbering of section to upper-case letters.
}

\usepackage[backend = biber,    % Recommended backend for sorting bibliography
            style = authoryear-comp,    % Close to the 'Harvard' referencing style
            urldate = long,     % Long: 24th Mar. 1997 | Short: 24/03/1997
            maxcitenames = 2,   % Number of authors in cite before replaced with 'Author#1 et al.'
            ]{biblatex}
\usepackage{subcaption}
\usepackage{minted} % Includes several programming languages and styles
\newcommand{\frontmatter}{
    \pagenumbering{roman}   % Setting page numbering to lower-case roman
}
\usepackage[british]{babel}     % Defining UK English as language. This will among other things ensure that dates are displayed as 24/03/1997 rather than 03/24/1997 in the bibliography.
\usepackage{import}     % Enable importing of sections
\usepackage{csquotes}   % Provides international handling of quote marks. Especially useful for bibliography management using BibLaTeX
\usepackage{hyperref}   % Hyper-references, possible to change color
\hypersetup{    % Color of hyper-references
    colorlinks,
    citecolor = black,
    filecolor = black,
    linkcolor = black,
    urlcolor = black
}
\usepackage{comment}    % Comment blocks of text
\usepackage{graphicx}   % Handle images
\usepackage{wrapfig}    % Wrap text around images
\usepackage{float}      % Force image location using "H"
\usepackage{url}
\newcommand{\HRule}{\rule{\linewidth}{0.5mm}}   % Ruler

\usepackage{siunitx}    % Enable SI units

\newcommand{\mainmatter}{
    \newpage
    \pagenumbering{arabic}  % Setting page numbering to normal integers
}

\usepackage{bm}     % Bold text in math mode
\usepackage{amsmath}    %  Math formulas and improved typographical quality of their output
\usepackage{amssymb}    % Extended symbol collection
\usepackage{amsthm}     % Helps define theorem-like structures

\usepackage[a4paper, total={150mm, 245mm,footskip = 14mm}]{geometry}
\usepackage{glossaries}
\usepackage{fancyhdr}
\usepackage{amssymb}
\usepackage[]{algorithm2e}

\usepackage{amsmath}
\makeglossaries

\newglossaryentry{Compmod}
{
    name=Compartmental model,
    description={A model which divides a population into different groups, where each group have different properties determinining how they interact with other groups}
}

\newglossaryentry{NetMod}
{
    name=Network model,
    description={In this context: A model consisting of vertices and edges, where the vertices represents the state of the model, while the edges represents the dynamic between them}
}
\newglossaryentry{EpiNetMod}
{
    name=Epidemiological Network Model,
    description={In this context: A network with vertices representing individuals/smaller population groups, and edges representing infection probabilities between the vertices}
}
\newglossaryentry{homoPop}
{
    name=Homogenous population,
    description={In this context: A population where the interacting compartments are close to equal size. (e.g the number of infected are close to the number of susceptible)}
}

\newglossaryentry{endmic}
{
    name= Endemic,
    description={The steady-state equilibrium after an epidemic}
}
\newglossaryentry{Rnot}
{
    name= Reproduction number,
    description={Number of average infections caused by an individual}
}

\newglossaryentry{MetaPop}
{
    name= Metapopulation,
    description={A population divided into smaller groups of individuals}
}

\newglossaryentry{MetaPopNet}
{
    name= Metapopulation network,
    description={In this context: A network model where vertices represents smaller groups of individuals, and the edges represents the dynamics between these groups}
}

\newglossaryentry{StochMod}
{
    name= Stochastic eifferential equations,
    description={In this context: Discrete differential equations where $\Delta x$ is sampled from probability distributions}
}
\newglossaryentry{ChainBinom}
{
    name= Chain Binomial differential equation,
    description={Differential equation where the next state is sampled from binomial probability distributions. In this context: Refers to the specific stochastic SIR/SEIR/SEIAR models used}
}

\newglossaryentry{BernoulliTri}
{
    name= Bernoulli trial,
    description={A binary outcome with a fixed probability}
}
\newglossaryentry{StrucProp}
{
    name= Structural property,
    description={Properties expressible purely in terms of the primitive mathematical theory of a model}
}
\newglossaryentry{Obs}
{
    name= Observability,
    description={A property describing how internal states of a system can be inferred from external measurements}
}

\newglossaryentry{Indent}
{
    name= Itentifiability,
    description={A property describing how system parameters can be inferred from external inputs/outputs}
}

\newglossaryentry{determinis}
{
    name= Deterministic system/model,
    description={A system/model with no randomness in the development of future states}
}

\newglossaryentry{Conjugate}
{
    name= Conjugate distributions,
    description={Distributions from the same family. (e.g. The Binomial and Beta distributions are in the same family, making the Beta distribution a conjugate prior for the Binomial)}
}

\newglossaryentry{Prior}
{
    name= Prior Distribution,
    description={A distribution expressing the belief of an uncertain quantity before evidence is taken into account}
}

\newglossaryentry{Posterior}
{
    name= Posterior distribution,
    description={A distribution expressing the belief of an uncertain quantity after evidence is taken into account}
}

\newglossaryentry{Dispersion}
{
    name= Dispersion,
    description={The statistical variability of a distribution describing how much the distribution is stretched/squeezed while retaining the same mean}
}

\newglossaryentry{PMF}
{
    name= PMF,
    description={Probability Mass Function, a discrete function}
}

\newglossaryentry{PDF}
{
    name= PDF,
    description={Probability Density Function, a continous function}
}

\newglossaryentry{Statmoment}
{
    name= Statistical moment,
    description={Quantitative measures related to the shape of a function. E.g. mean, variance and skewness of a distribution}
}

\newglossaryentry{ideal}
{
    name= Ideal,
    description={A subset of elements which forms an additive group. (If x and y are contained in the ideal, $xy$ is also contained)}
}
\newglossaryentry{CharSet}
{
    name= Characteristic set,
    description={A minimal set of differential polynomials generating the same differential ideal as an arbitrary set of the same polynomials}
}
\newglossaryentry{pseudorem}
{
    name= Pseudoremainder,
    description={The remainder after a polynomial division}
}

\newglossaryentry{groebner}
{
    name= Gr\"obner bases,
    description={A special generating set for an ideal. Can be considered a generalization of gaussian elimination, applied on polynomials instead of linear equations}
}
\newglossaryentry{autoreduced}
{
    name= Autoreduced,
    description={In this context: An equation from a set is autoreduced if it cannot be reduced by the other equations within that set}
}
\newglossaryentry{diffField}
{
    name= Differential field,
    description={A differentiable set where addition, subtraction, multiplication and division is defined and behave corresponding to the same operation on rational and real numbers}
}
\newglossaryentry{degeneracy}
{
    name= Degeneracy,
    description={In this context: The limiting case where a class of particles changes their nature to belong to a simpler class. When particles explore approximately the same state space, they are degenerate compared to particles with widely different trajectories}
}
\newglossaryentry{markovModel}
{
    name= Markov model,
    description={A stochastic model where the future state only depends on the present and not previous events}
}
\newglossaryentry{reduction}
{
    name= Reduction,
    description={Rewriting a mathematical expression to a simler form}
}
\author{ }
\date{ }
\begin{document}

% Inserting title page
% Inspired by title template from ShareLaTeX Learn; Gubert Farnsworth & John Doe
% Edited by Jon Arnt Kårstad, NTNU IMT

\begin{titlepage}
\vbox{ }
\vbox{ }
\begin{center}
% Upper part of the page
\includegraphics[width=0.40\textwidth]{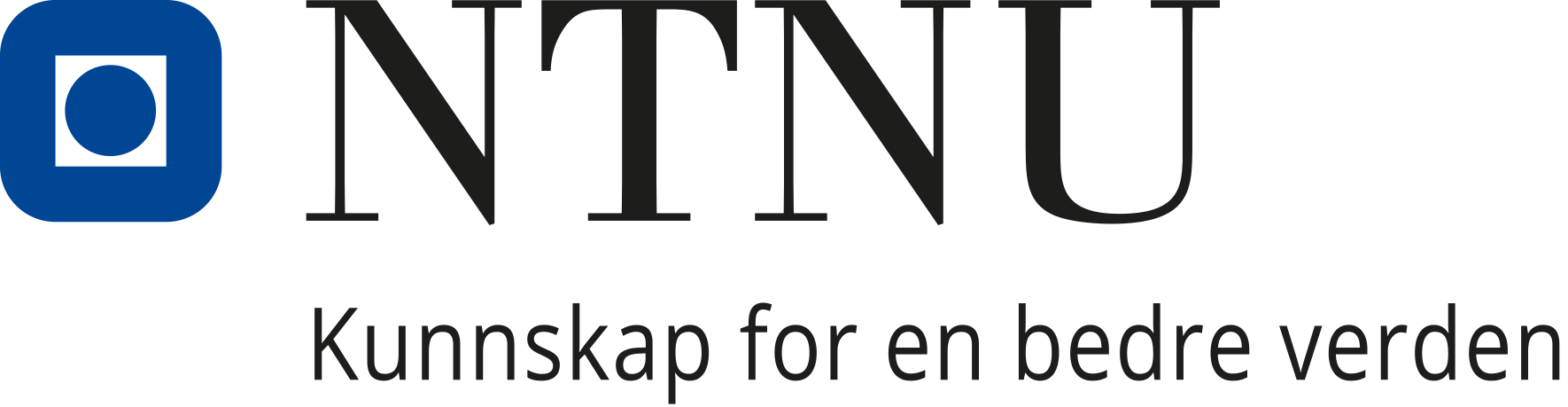}\\[1cm]
\textsc{\LARGE Department of Engineering Cybernetics}\\[1.5cm]
\textsc{\Large TTK4551 - Specialization Project }\\[0.5cm]
\vbox{ }

% Title
\HRule \\[0.4cm]
{\huge \bfseries Identifiability, Observability, Uncertainty and Bayesian System Identification of Epidemiological Models}\\[0.4cm]
\HRule \\[1.5cm]

% Author
\large
Jonas Hjulstad
\vfill

% Bottom of the page
{\large May, 2021}
\end{center}
\end{titlepage}

\thispagestyle{plain}
\begin{center}
In this project, identifiability, observability and uncertainty properties of the deterministic and Chain Binomial stochastic SIR, SEIR and SEIAR epidemiological models are studied. Techniques for modeling overdispersion are investigated and used to compare simulated trajectories for moderately sized, homogenous populations. With the chosen model parameters overdispersion was found to have small impact, but larger impact on smaller populations and simulations closer to the initial outbreak of an epidemic. 

Using a software tool for model identifiability and observability (DAISY[\cite{ref_DAISY}]), the deterministic SIR and SEIR models was found to be structurally identifiable and observable under mild conditions, while SEIAR in general remains structurally unidentifiable and unobservable.

Sequential Monte Carlo and Markov Chain Monte Carlo methods were implemented in a custom C++ library and applied to stochastic SIR, SEIR and SEIAR models in order to generate parameter distributions. With the chosen model parameters overdispersion was found to have a small impact on parameter distributions for SIR and SEIR models. For SEIAR, the algorithm did not converge around the true parameters of the deterministic model. The custom C++ library was found to be computationally efficient, and is very likely to be used in future projects.
\end{center}
\newpage
% Defining front matter settings (Norsk: innstillinger for forord m.m.)
\frontmatter

% Inserting table of contents
\tableofcontents

% Inserting list of figures & list of tables
\listoffigures
\listoftables
\clearpage
\glsaddall
\printglossaries
% Defining main matter settings (Norsk: innstillinger for hoveddelen av teksten)
\mainmatter

% Introduction explaining this LaTeX-template
\section{Introduction}

\subsection{Background}
Variants of deterministic, compartmental models have been developed to
model epidemics for over a century, starting with bilinear dynamics proposed in 1906 [\cite{hamer1906milroy}], further developed to a three-compartment model called the Kermack-Mckendrick model (SIR) in 1927 [\cite{kermack1927contribution}]. The dynamics of these models are well understood and favorable for analytical purposes, but the models are limited to capturing the dynamics of large, homogenous populations. 

Past epidemics in the 20th century have re-ignited the interest for epidemic modeling in general, including the use of network models and stochastic designs, in order to capture the dynamics of heterogenous populations. In the shortcomings for deterministic models in small populations or initial/endemic stages of a disease, stochastic models manage to capture the events, but the number of possible outcomes for these models may be large. Network models have also been used for a long time in epidemiology, originating from the study of graphs by Erdös and Rényi [\cite{renyi1960evolution}]. These models makes it possible to specify the structure within the population, but falls short on the analytical properties, making it harder to determine and relate the reproduction number $\mathscr{R}_0$ and final size of the epidemic. 

The advantages of network models have been adressed by NTNUs COVID-19 modelling taskforce   [\cite{ref:NTNU_COVID}], a model that introduce a metapopulation model fitted to capture social contact in schools, workplaces, nursing homes and hospitals in Trondheim and Oslo. 

Stochastic differential equations were introduced earlier by Reed and Frost in 1928 [\cite{reed_frost}], which introduced random binomially distributed infections to compartmental models. These fall under the category of Chain-Binomial models, which model the chance of escaping infection and other transitions as a series of Bernoulli trials. These models have better analytical properties than network models, but they do not use spatial information, which makes the overall uncertainty of the model much larger. 

\subsection{Problem Statement}

The objective of this project is to study structural identifiability and state observability of simple, compartmental epidemiological models, specifically SIR/SEIR/SEIAR models.  In addition, the representation of uncertainty in simple compartmental models should be addressed, and how the uncertainty in simple compartmental models can be represented by appropriately selecting parameter distributions in the corresponding stochastic version of the same compartmental models.  The focus should be on moderately sized populations where the assumptions underlying a deterministic approximation do not hold.

The project should:
\begin{itemize}
    \item Provide an introduction to the probability distributions used to describe uncertainty in epidemiological models
    \item  Provide an introduction to SIR/SEIR/SEIAR-type compartmental epidemiological models.
    \item Analyze structural identifiability and observability in the deterministic compartmental models.
    \item Study how stochastic Chain-Binomial models can be used to generate parameter distributions for compartmental models.  To this end, Sequential Monte Carlo- and Markov Chain Monte Carlo-based methods should be studied.
    \item Create a software library with an efficient SMC/MCMC-sampler capable of estimating parameter distriutions for the introduced models. 

\end{itemize}

\subsection{Motivation}
There are several reasons for investigating the aforementioned topics. Relating deterministic to stochastic models is an important step towards connecting the models to true epidemics, which are very stochastic by nature. This is also a first step towards relating the deterministic model to network models, for which there exist ways of assessing model parameters on a smaller scale, which as a consequence results in overparametrized, complex models. Finding reasonable approximations can reduce the computational expense of simulations, and make it possible to retrieve locally valid, analytical solutions which makes it easier to understand the course of the epidemic itself.

Another reason is controlling the outcome of epidemics. Numerical optimal control is a well established field for deterministic models, which makes the deterministic approximations very attractive as long as they remain valid. Considering that political decisions in an epidemic may have lethal consequences, it is important that robust control strategies are implemented, which in turn requires the validity of model approximations to be well understood.

\subsection{Software}
Python was initially chosen as programming language for this project, but computational efficiency moved parts of the project over to C++. Compatibility issues and challenges with the building of C++ libraries and Python dependencies made Ubuntu a more attractive operating system.

\begin{table}[H]
    \centering
    \begin{tabular}{|c|c|c|c|c|c|}
    \hline
        \textbf{Software} & \textbf{Version} & \textbf{Description} & \textbf{Software} & \textbf{Version} & \textbf{Description}\\\hline
        \textbf{Python} & 3.8 &Interpreter &Matplotlib & 3.4.1 & Plotting\\
        PyMC3 & 3.11.2 & SMC/MCMC& Theano-PyMC & 1.1.2 & Optimized Compilation\\
        Scipy & 1.6.3 & Sampling and Integration & & &\\\hline
        \textbf{C++} & ISO C++14 & Language Standard & g++  & 9.3.0 &Compiler\\
        CMake & 3.16.3 & Build automation & GSL & 2.5 & Standard Library\\\hline
        \textbf{Ubuntu} & 20.04.02 LTS & Operating System & & &\\\hline

    \end{tabular}
    \caption{Software and libraries used}
    \label{tab:Software_used}
\end{table}

% Example section added from an external tex-file, here located in ./Sections/
\section{Probability Distributions and Dispersion}
\label{ch:Dispersion}
Epidemics are in general well approximated by deterministic differential equations when the susceptible and infected groups of a large population are homogenously distributed and of equal size. This does not hold at the time of outbreak and close to the endemic timepoint, where the number of infected is smaller. The deterministic models also fail to capture the dynamics of smaller populations.

As an alternative to deterministic dynamics, it is possible to model infections, exposure and recovery with probabilities. The distributions necessary to model these effects are introduced in this section. In some cases of modeling, the variance of a distribution may be expected to be higher than the base variance of the candidate distribution. In this case the model is overdispersed, which will be shown to be implementable by combining distributions.
\subsection{Binomial Distribution}
The Binomial distribution captures the probability of $k$ successful Bernoulli trials (with probability $p$) occuring in $n$ attempts. Its probability mass function (PMF) is given by:
\begin{equation}
    P_{PMF}(X=k | n, p) = \binom{n}{k}p^k(1-p)^{n-k} \triangleq \text{Bin}(k,n,p)
\end{equation}
The distribution has an expected value of $np$ with variance $np(1-p)$. The Binomial distribution is useful for modeling infections and recoveries when they are considered to be independent trials with equal probability. The variance of the distribution converges to the mean when the number of trials is large, and the probability is small ($np(1-p) \approx np$ when $n >> p$).

\subsection{The Poisson Distribution}
The Poisson distribution captures the probability of a number of events k to occur in a time interval $t$ with event rate $r$. Its PMF is parametrized by the expected number of events $\lambda = rt$. 

\begin{equation}
    P_{PMF}(X=k | \lambda) = \frac{\lambda^k e^{-\lambda}}{k!} \triangleq \text{Po}(k, \lambda)
\end{equation}

The distribution models systems with a large number of possible, rare events well, but fails to represent where probabilities are higher or the number of events lower.

\subsection{Gamma Distribution}
The Gamma distribution is parametrized with $\alpha$ and $\gamma$, giving the following Probability Density Function (PDF):
\begin{equation}
    P_{PDF}(X=k|\alpha, \beta) = \frac{\beta^\alpha k^{\alpha - 1}e^{-\beta k}}{\Gamma(\alpha)} \triangleq \text{Gamma}(k, \alpha, \beta)
\end{equation}
Where $\Gamma(\alpha)$ is the gammma function. The distribution has important relations to the exponential and normal distributions, and is a popular choice to combine with other distributions to get desired properties of moments.
\subsection{Negative Binomial Distribution}
\label{ch:NegBin}
\iffalse
The model can simulate undetected individuals by adding an observation process onto the infected state. Given a detection probability $p_obs$ and a maximum number of detection failures $s$, the observation process can be assumed to be a Negative Binomial distribution.

\begin{equation}
    Y_t \sim \text{NegBin}(\mu = p_{obs}I_t, \mbox{size}=s)
\end{equation}
\fi

The Negative Binomial distribution can be seen as a gamma-distributed variable drawn from a Poisson distribution, where the PMF is given by:
\begin{equation}
    P_{PMF}(X=k | \lambda, r) = \frac{\lambda^k}{k!}\frac{\Gamma(\lambda + k)}{\Gamma(r)(r + \lambda)^k}\frac{1}{(1+r)^r} \triangleq \text{NegBin}(k, \lambda, r)
\end{equation}
The Poisson-Gamma relationship to the Binomial distribution is seen by marginalizing the joint distribution of $\lambda, k$ with respect to $\lambda$:
\begin{equation}
    \text{NegBin}(X=k | \lambda, r) = \int_0^\infty \text{Poisson}(k, \lambda)\text{Gamma}(X=\lambda, \alpha = r, \beta = \frac{1-p}{p}) d\lambda
\end{equation}

Where $r$ is the number of failures to be drawn from a set of $k$ bernoulli trials with success-probability $p$ and mean $\lambda = \frac{pr}{1 - p}$. When the number of failures to draw approaches infinity, the Negative Binomial distribution converges towards the Poisson distribution:
\begin{equation}
    \lim_{r\rightarrow \infty} = \frac{\lambda^k}{k!}\frac{1}{1}\frac{1}{e^{\lambda}}
\end{equation}

For finite values of $r$ the mean and variance of the distribution is shifted to represent overdispersion. This can also be seen in the variance of the distribution:
\begin{equation}
    \lim_{r\rightarrow \infty}E[(x-\Bar{x})^T(x-\Bar{x})] = \lim_{r\rightarrow \infty} \lambda(1 + \frac{\lambda}{r}) = \lambda
\end{equation}
Setting $\nu = \frac{\lambda}{r}$ makes it possible to scale the distribution variance with $\nu$ proportionally to the Poisson base variance $\lambda$.
\subsection{Beta Distribution}
Similar to the Gamma distribution the Beta distribution is another choice to adjust moments of distributions. It can be considered a function of two Gamma-distributed variables:
\begin{align}
\begin{split}
    \frac{X}{X+Y}&\sim \text{Beta}(\alpha, \beta) \\
    X \sim \text{Gamma}(\alpha, 1) &
    ,\quad Y \sim \text{Gamma}(\beta, 1)
\end{split}
\end{align}
This results in the following PDF:
\begin{equation}
    P_{PDF}(X=k | \alpha, \beta) = \frac{1}{B(\alpha, \beta)}k^{\alpha-1}(1-k)^{\beta-1} \triangleq \text{Beta}(\alpha, \beta)
\end{equation}
Where $B(\alpha, \beta)$ is the Beta function.
\subsection{Beta-Binomial Distribution}
The Beta-Binomial distribution can be seen as a Binomial distribution where $p$ is drawn randomly from a Beta-distribution with the following PMF:
\begin{equation}
    P_{PMF}(X=k|\alpha, \beta, n) = \binom{n}{k}\frac{B(k + \alpha, n-k+\beta)}{B(\alpha, \beta)} \triangleq \text{BetaBin}(\alpha, \beta, n)
    \label{eq:betabinPMF}
\end{equation}
The first moments are given by equations \ref{eq:betabinMoment}, \ref{eq:betabinVar}.
\begin{align}
    E[X] &= \frac{n\alpha}{\alpha + \beta} \label{eq:betabinMoment}\\
    E[(X-\Bar{x})^T(X-\Bar{x})] &= \frac{n\alpha\beta(\alpha + \beta + n)}{(\alpha+\beta)^2(\alpha + \beta + 1)}\label{eq:betabinVar}
\end{align}
This distribution will only be used for overdispersion of the Binomial distribution, but it should be noted that a relation to the Poisson distribution can be derived through marginalization much in the same way as the Negative Binomial distribution.

The relationship between the Beta, Binomial and Beta-Binomial distributions is seen by marginalizing the joint distribution of $p, k$ with respect to $p$. 

\begin{equation}
    \text{BetaBin}(X=k | p, n) = \int_0^1 \text{Bin}(k| n,p )\text{Beta}(p |\alpha, \beta) dp
    \label{eq:marg_betabin}
\end{equation}

By assigning $\mu = \frac{\alpha}{\alpha + \beta}$ and $\gamma = \frac{1}{\alpha + \beta + 1}$ the first moments can be reparametrized.

\begin{align}
    E[X] &= n\mu\label{eq:reparametrized_mu_betabin}\\
E[(X-\Bar{x})^T(X-\Bar{x})] &= n\mu(1-\mu)(1+(n-1)\gamma)\label{eq:reparametrized_sig_betabin}
\end{align}
$\gamma \in (0,1)$ is the dispersion parameter. When $\gamma \rightarrow 0$ the distribution variance converges to the variance of the Binomial distribution. Setting $\gamma = \frac{\nu}{n-1}$ makes it possible to scale the distribution variance with $\nu$ proportionally to the Binomial base variance $n\mu(1-\mu)$.

\subsection{Comparison of Distributions and Limitations}
\begin{figure}[H]
    \centering
    \includegraphics[width=.8\textwidth]{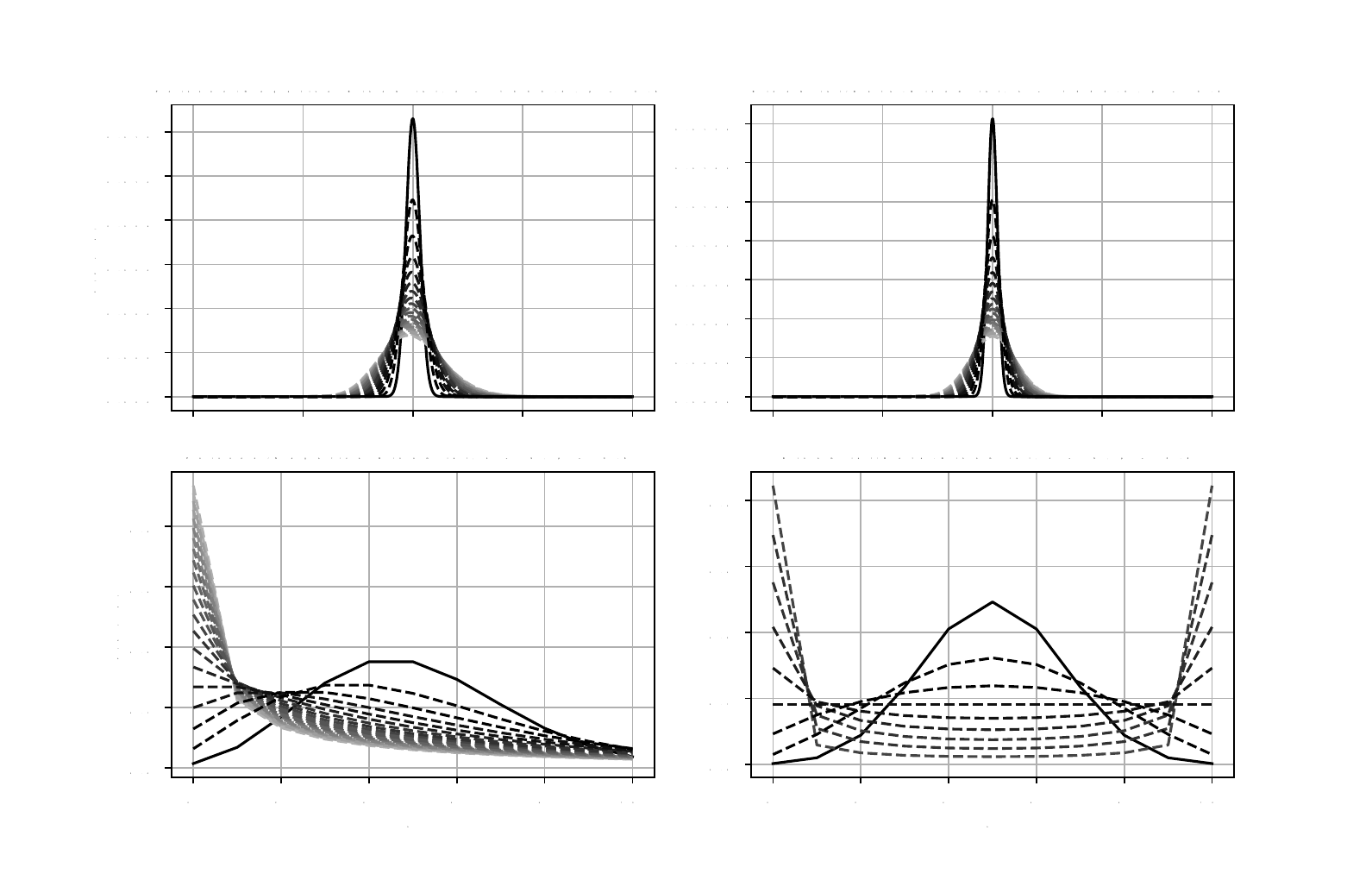}
    \caption{Comparison of overdispersion using PMFs for the Poisson, Negative Binomial, Binomial and Beta Binomial distributions. The dispersion is linearly spaced with additional $\nu = [1, \dots, 8]$ standard deviations compared to the base distributions. }
    \label{fig:Poisson_Binomial_Comparison}
\end{figure}

Given a small probability $p$ and large sample size $n$, the Poisson distribution approximates the Binomial distribution well, even with overdispersion. Figure \ref{fig:Poisson_Binomial_Comparison} illustrates differences in dispersion for the approximations.

The Poisson distribution keeps the same shape as the Binomial distribution even when the approximation is bad, but the Negative Binomial and Beta Binomial distributions become very different in this case. It is therefore very important to ensure that Poisson approximations remain valid for the whole trajectory if they are to be used.

\section{Epidemiological Models}
\label{ch:EpidModels}
The stochastic models introduced are based on terms from a demographic, stochastic SEIR-model [\cite{ref:Sochastic_SEIR}]. The demographic terms excluded in order to retain the property of a constant total population. Poisson and Binomial distributions are used to present the dynamics, which respectively can be replaced with negative Binomial and Beta-Binomial distributions to model overdispersion (Section \ref{ch:Dispersion}).

Deterministic and corresponding Chain-Binomial models are introduced for the classic SIR, SEIR and SEIAR compartmental models, followed by trajectory simulations to compare the deterministic and stochastic models for small populations. The initial state is set to a timepoint after the initial outbreak, which is captured by the following deterministic parameters:
\begin{table}[H]
    \centering
    \begin{tabular}{|c|c|c|c|c|c|}
    \hline
        $\Delta t$ &  $1[\text{Day}]$ &
         $T_{\text{span}}$ & $[0,201] [\text{Days}]$ & $S_0$ &$9000/900\text{(SIR)}$\\\hline
         $\mathscr{R}_0$ & $1.2$ & $\alpha$ & $0.11$ & $I_0$  & $1000/100\text{(SIR)}$ \\\hline
         $\beta$ & $0.13$
          & $\gamma$& $0.33$ & $R_0$&$0$\\\hline
         $\mu$ & 0.11 &p & 0.5& $N_{pop}$ & $10000/1000 \text{(SIR)}$\\\hline
         
    \end{tabular}
    \caption{Deterministic parameters used for epidemiological models}
    \label{tab:Determ_param}
\end{table}

\subsection{SIR Model}
The SIR discrete model is minimal in its number of compartments for modeling epidemics without periodicity. The average contact rate per person $\beta$ and recovery rate $\alpha$ are the only parameters required to describe an asymptotically stable epidemic. 

\textbf{Deterministic:}
\begin{align}
    \begin{split}
        \dot{S} &= S - \beta \frac{SI}{N_{pop}}\\
        \dot{I} &= I + \beta \frac{SI}{N_{pop}} - \alpha I\\
        \dot{R} &= R + \alpha I
    \end{split}
    \label{eq:SIR_model}
\end{align}
The Susceptible ($N$), Infected ($I$) and Recovered ($R$) population groups adds up to total population $N$ at all timesteps. Given an constant population $N_{pop}$ the third equation can be omitted and retrieved post-simulation as $R(t) = N_{pop} - S(t) - I(t)$.

\textbf{Stochastic:}
\begin{align}
\begin{split}
    \Delta S &= - \text{Po}(S_k \cdot p_I(t))\\
    \Delta I &= \text{Po}(S_k \cdot p_I(t)) - \text{Bin}(I_k, p_R)\\
    \Delta R &= \text{Bin}(I_k, p_R)
    \label{eq:Stochastic_SIR}
\end{split}
\end{align}
Where $p_I(t) = 1-\text{exp}[-\beta c(t)\frac{I(t)}{N_{pop}}\Delta t]$, and $\beta, c(t)$ are the transmission probability and contact rate respectively. Fixed probabilities can be fitted to the deterministic model, yielding $p_R = 1-\text{exp}(-\alpha \Delta t)$.

\textbf{Note: }With some abuse of notation, the probability-distributed variables are drawn once every iteration (e.g. $\text{Po}(S_k\cdot p_I(t))$ is the same sample for $\Delta S, \Delta I$ in equation \ref{eq:Stochastic_SIR}, not two samples from the same distribution).

\subsubsection{Simulated Trajectories}
\begin{figure}[H]
    \centering
    \includegraphics[width=\textwidth]{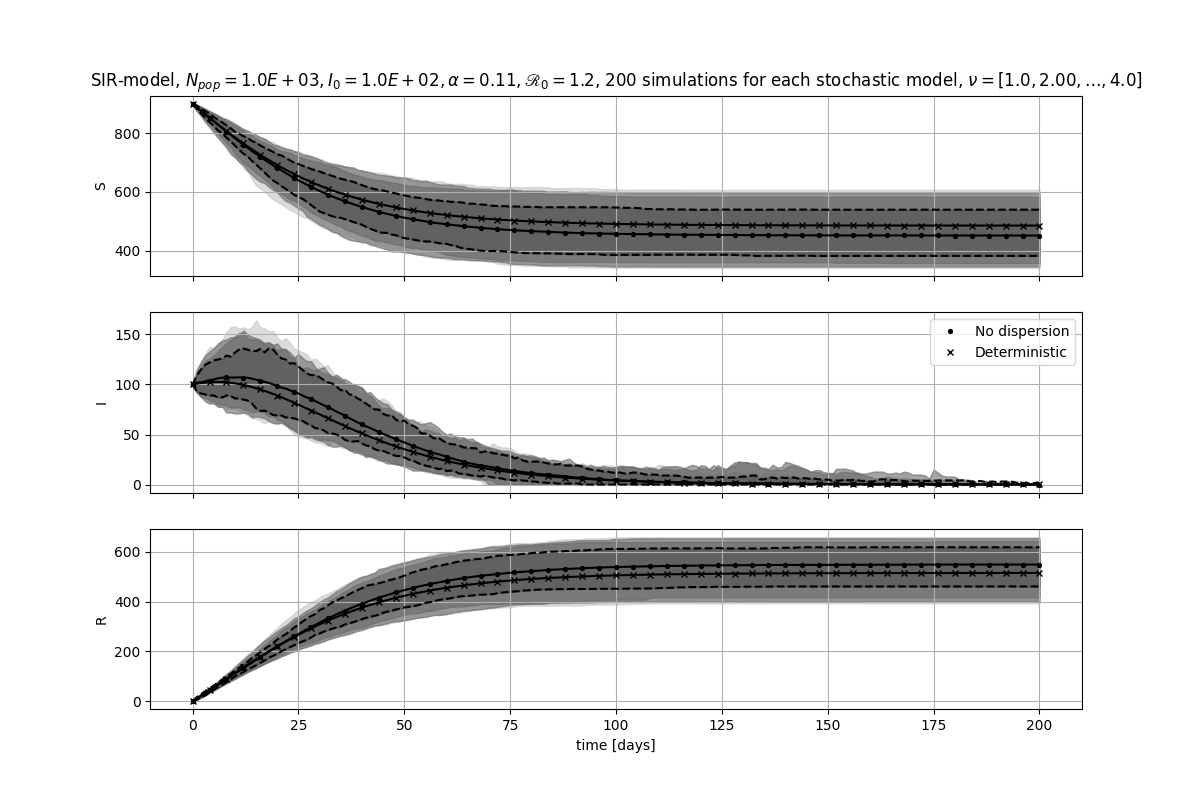}
    \caption{Comparison of dispersed trajectories for SIR-model ($90$-percentiles from both sides of the  stochastic simulation mean). Increasing $\nu$ gradually fades in to gray. 90-percentiles for undispersed trajectory is shown in dotted lines.}
    \label{fig:SIR_disperse_comp}
\end{figure}

\subsection{SEIR Model}
An intermediate stage can be added to the SIR model in order to model incubation period. This compartment is very relevant for the current pandemic, considering the expected average incubation period of $5.1$ days[\cite{ref:incub_time}] for COVID-19.

\textbf{Deterministic:}
\begin{align}
\begin{split}
    \dot{S} &= -\beta \frac{SI}{N_{pop}}\\
    \dot{E} &= \beta \frac{SI}{N_{pop}} - \gamma E\\
    \dot{I} &= \gamma E - \alpha I\\
    \dot{R} &= \alpha I
\end{split}
\end{align}

\textbf{Stochastic:}
\begin{align}
\begin{split}
    \Delta S &= - \text{Po}(S_k \cdot p_E(t))\\
    \Delta E &= \text{Po}(S_k \cdot p_E(t)) - \text{Bin}(E_k, p_I)\\
    \Delta I &=  \text{Bin}(E_k, p_I) - \text{Bin}(I_k, p_R)\\
    \Delta R &= \text{Bin}(I_k, p_R)
\end{split}
\end{align}
Where $p_E(t) = 1-\text{exp}[-\beta c(t)\frac{I(t)}{N_{pop}}\Delta t]$ and $p_I = 1-\text{exp}(-\gamma \Delta t)$.

\subsubsection{Simulated Trajectories}
\begin{figure}[H]
    \centering
    \includegraphics[width=\textwidth]{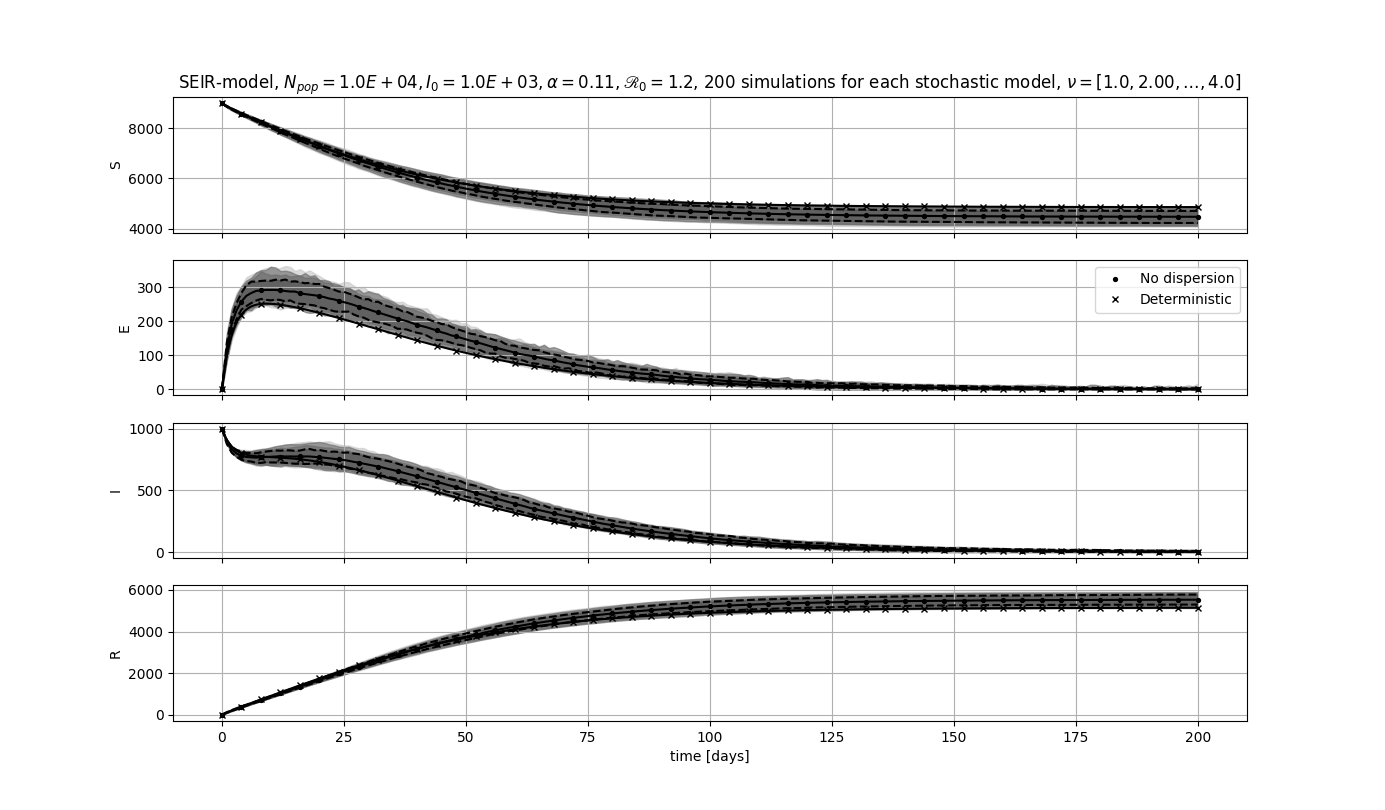}
    \caption{Comparison of dispersed trajectories for SEIR-model ($90$-percentiles from both sides of the  stochastic simulation mean). Increasing $\nu$ gradually fades in to gray. 90-percentiles for undispersed trajectory is shown in dotted lines.}
    \label{fig:SEIR_disperse_comp}
\end{figure}

\subsection{SEIAR Model}
Testing unavailability and high estimated asymptomatic rates for COVID-19[\cite{ref:High_Asymptomatic_US}], indicates that there should be an alternative compartment for some infected individuals. A SIR/SEIR model fitted on observed infections will not be able to account for undetected infections, which is why an asymptomatic group is introduced.

\textbf{Deterministic:}
\begin{align}
\begin{split}
    \dot{S} &= -\beta \frac{S(I+A)}{N_{pop}}\\
    \dot{E} &= \frac{S(I+A)}{N_{pop}} - \gamma E\\
    \dot{I} &= \gamma p E - \alpha I\\
    \dot{A} &= \gamma(1-p)E -\mu A\\
    \dot{R} &= \alpha I + \mu A
\end{split}
\end{align}

The fraction of exposed turning infectious is described by $p$, and the recovery rate of the asymptomatic group is described by $\mu$. 

\textbf{Stochastic:}
\begin{align}
\begin{split}
    \Delta S &= - \text{Po}(S_k \cdot p_E(t))\\
    \Delta E &= \text{Po}(S_k \cdot p_E(t)) - \text{Bin}(E_k, p_I) - \text{Bin}(E_k, p_A)\\
    \Delta I &=  \text{Bin}(E_k, p_I) - \text{Bin}(I_k, p_{R,I})\\
    \Delta A &=  \text{Bin}(E_k, p_A) - \text{Bin}(A_k, p_{R,A})\\
    \Delta R &= \text{Bin}(I_k, p_{R,I}) + \text{Bin}(A_k, p_{R, A})
\end{split}
\end{align}

Where $p_E(t) = 1-\text{exp}[-\beta c(t) \frac{I(t) + A(t)}{N_{pop}}\Delta t]$, $p_I = 1-\text{exp}(-p\gamma \Delta t)$ and $p_A = 1-\text{exp}(-(1-p)\gamma \Delta t)$. $p_{A,R}, p_{I,R}$ follows the same pattern.
\subsubsection{Simulated Trajectories}
\begin{figure}[H]
    \centering
    \includegraphics[width=\textwidth]{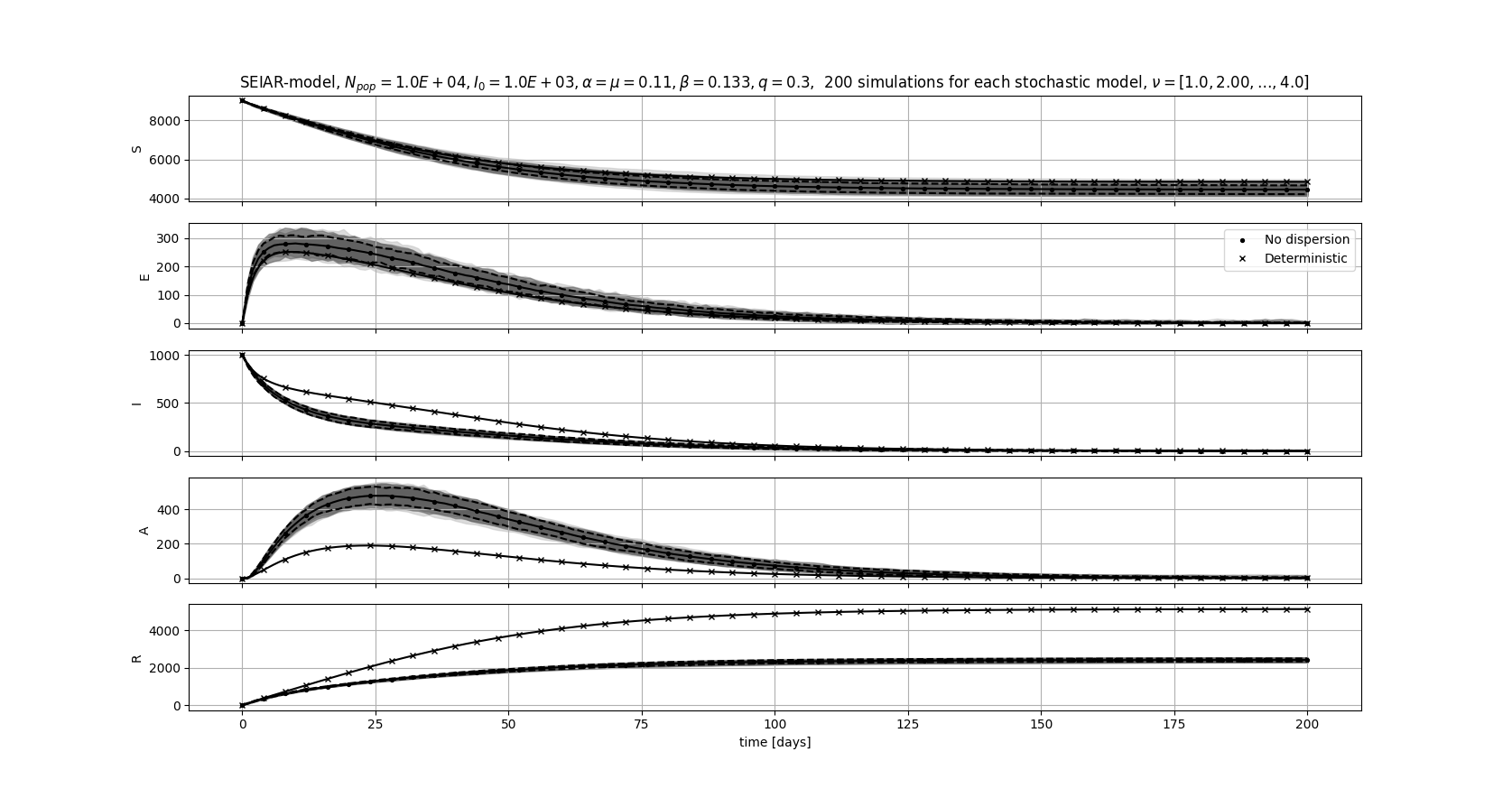}
    \caption{Comparison of dispersed trajectories for SEIAR-model ($90$-percentiles from both sides of the  stochastic simulation mean). Increasing $\nu$ gradually fades in to gray. 90-percentiles for undispersed trajectory is shown in dotted lines.}
    \label{fig:SEIAR_disperse_comp}
\end{figure}
\iffalse

\subsection{Reed-Frost Model}
The Reed-Frost model is a discrete, stochastic version of the SIR model. Instead of a constant flow of individuals from the susceptible to infected group, there is a probability of escaping infection given by the previous number of infected individuals ($(1 - p)^{I_{t-1}}$). Thus, the total number of new infected can be determined by a binomial distribution with the remaining probability:
\begin{align}
    \begin{split}
        I_t &\sim \text{Bin}(S_{t-1}, 1 - (1-p)^{I_{t-1}})\\
        S_t &= S_{t-1} - I_t
    \end{split}
\end{align}

\fi

\subsection{A Simulation of the Initial Outbreak}
With a smaller number of initial infected, the interval of outcomes becomes much larger. 
\begin{figure}[H]
    \centering
    \includegraphics[width=.8\textwidth]{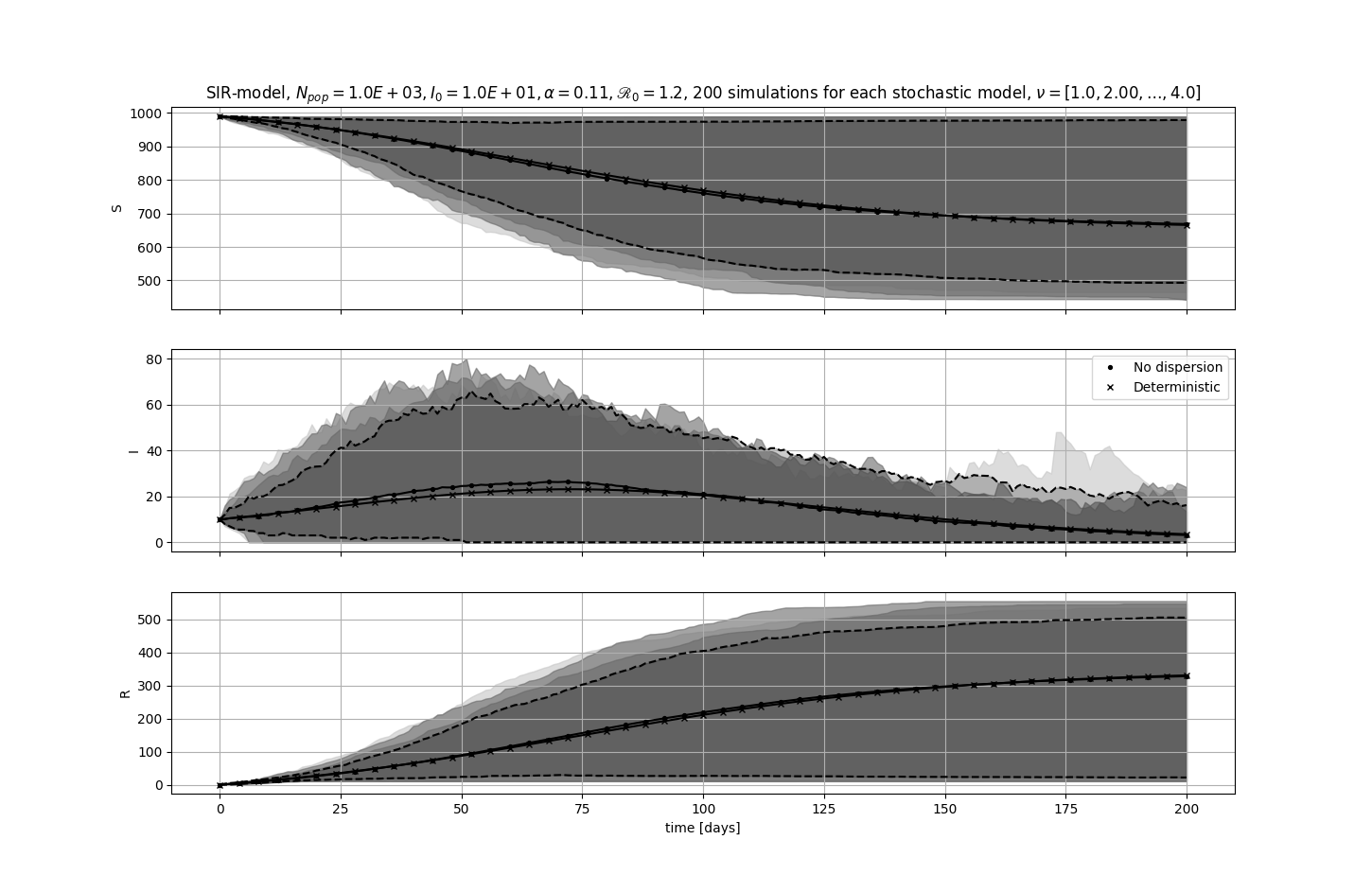}
    \caption{Comparison of dispersed trajectories for SIR-model ($90$-percentiles from both sides of the  stochastic simulation mean). Increasing $\nu$ gradually fades in to gray. 90-percentiles for undispersed trajectory is shown in dotted lines.}
    \label{fig:Initial_Outbreak_SIR}
\end{figure}

\subsection{Remarks on the Simulations}
The parameters used was specifically chosen to yield trajectories sensitive to distribution variance, where the average mean trajectory still is relatable to the deterministic one.

Stochastic SEIR/SEIAR has one extra binomial term in the connection between the susceptible and the infected group, which results in an overall variance increase in the resulting trajectories. In order to illustrate trajectories on a comparable scale, the total population in SEIR/SEIAR-simuations is larger.

For the total population size $N_{pop}=10000/1000(SIR)$ with $\Delta t = 1[\text{Day}]$ the mean stochastic trajectory is not approximately equal to the deterministic. This changes when the exponent in the binomial probabilities ($p_I, p_E, p_A$, ..) converges towards $0$, which is achieved by increasing $N_{pop}$ or decreasing $\Delta t$. 

The asymmetric 90-percentiles in the illustrations are not approximately equal to the true confidence intervals. The methods implemented to obtain true confidence intervals in Python ended up becoming too computationally expensive to evaluate. If more accurate intervals are needed in the future, a reimplementation in C++ could be considered.

The dispersion parameter $\nu$ was expected to have a larger impact. Increasing the variance by a factor of 5 ($\nu = 4$) was expected to drastically impact the resulting trajectories, but they remain comparatively close for this population size. Further reducing the population size enhances the effect of the dispersion, but also separates the stochastic mean from the deterministic trajectory. The 90-percentiles may be decieving in this case, a better approximation of the true confidence intervals may be needed.

\subsection{Discussion}
The small effect of overdispersion on the trajectories for populations with homogenous initial populations is promising for the deterministic approximation. If the deterministic approximation remains valid for parts of the epidemic, it becomes possible to consider applying numerical optimal control directly on the model equations, which is a much simpler approach than the robust alternatives needed for stochastic models.

However, when trajectories are simulated from the initial stage of an outbreak (figure \ref{fig:Initial_Outbreak_SIR}) the 90-percentiles cover a much larger region of the state space. Performing optimal control on a deterministic approximation could in this case result in solutions which underestimates the infectious spread, and should therefore be avoided. The space of possible outcomes is very large for Chain Binomials under these conditions, which raises the need for individual based modeling. 

The different regions of validity shows that an understanding of a true epidemic may require the use of multiple models, where each model is valid in different regions. It is reasonable to assume that the initial outbreak is better modeled with spatial information and a more detailed description of each individuals dynamic.

The 90 percentiles produced with no dispersion covers a significant region. The effect of underdispersion should be investigated to address this.

\section{Identifiability}
Verifying the performance of estimators is challenging with low availability of data. A previous approach to optimal control of the COVID-19 pandemic ([\cite{Africa_study}]) have addressed this by bounding the parameter space to reasonable limits before performing the identification. Even with a restricted parameter space, it is still uncertain if the data is informative enough to distinguish the parameter values, or if the resulting model will be overfitted. With reasonable parameter ranges, this can still be solved with robust control strategies ([\cite{kohler2020robust}]), but this does not solve the validation-issues of the models.

One solution ([\cite{Brazil}] have split data into training/validation datasets to verify the performance of identified parameters. While verifying the model on currently available data makes is a good practice for confirming how well it performs in a local region, it goes not guarantee good prediction performance for the future. 

Before performing parameter identification, it is important to
determine if some parameters even are identifiable at all. This is done by evaluating the structural idenitifiability, which in the following sections will be done algebraically by hand and with specifically designed software.

\subsection{Differential Algebra for Identifiability of Systems}
Differential Algebra for Identifiability of SYstems (DAISY) [\cite{ref_DAISY}] is a software tool used to assess global/local identifiability of parameters in polynomial/rational models. It is based on the general-purpose algebra system REDUCE [\cite{ref_REDUCE}], and can be used to determine the parameter identifiability in some of the presented epidemiological models. 
\subsubsection{Ranking}
DAISY uses an algorithm which requires a variable ranking of the inputs, outputs and internal states of the subject system and their derivatives. The ranking does not impact the outcome of the algorithm, but may reduce the number of divisions needed to produce results. 
\begin{align}
    u_1 < u_2 &< \dots < y_1 < y_2 < \dots \dot{u}_1 < \dot{u}_2 < \dots < \Ddot{u}_1 < \Ddot{u}_2\\ < \dots < &\dot{y}_1 < \dot{y}_2 < \dots < \Ddot{y}_1 < \Ddot{y}_2 < \dot{x}_1 < \dot{x}_2 < \dots\nonumber
\end{align}
For a polynomial $A_i$ the term containing the highest ranked variable (or highest ranked variable product, e.g. $\dot{x}_1y_1 < \dot{x}_1\dot{y}_1$) is called the \textit{leader} term.  

\subsubsection{Pseudodivision Algorithm}
\label{ch:pseudodiv}
To determine the uniqueness of parameter solutions it is desired to find the \textit{characteristic set} of the subject system. The characteristic set is a minimal set of differential polynomials which generates the same differential \textit{ideal} as an arbitrary set of the same polynomials. A polynomial $A_i$ is said to be reduced wrt. $A_j$ if it does not contain any algebraic degreee of $A_j$'s leader term or its derivatives. With the previous definitions, \textit{Ritt's pseudodivision algorithm} [\cite{ref:rittPseudo}] can be summarized in three steps:

\begin{enumerate}
    \item If polynomial $A_i$ contains the $k$th derivative of $A_j$'s leader term it is differentiated $k$ times, resulting in a polynomial with same leader term as $A_j$.
    \item Multiply $A_i$ with the coefficient of the leader term, then divide the result by $A_i$'s $k$th derivative. Let $R$ be the \textit{pseudoremainder} from the division. 
    \item Replace $A_i$ with $R$ and the $k$th derivative with the $(k-1)$th of $A_j$ and repeat the procedure until $R$ is reduced with respect to the $0$th derivative of $A_j$.
\end{enumerate}
Pseudodivisions are applied to all pairs of differential polynomials (system equations) until they are all reduced with respect to each other. The resulting set is called an \textit{autoreduced set}, which in its lowest rank is the characteristic set of the system.

\subsubsection{Parameter Solutions from the Characteristic Set}

The input-output relation of the system is fully represented in one of the polynomials of the characteristic set. The coefficients from the input-output polynomial can be extracted and fixed to a set of values. This yields a set of solvable, nonlinear equations which DAISY solves by finding the Gr\"obner basis using Buchberger's algorithm (see Appendix A).

The Gr\"obner basis for polynomials serve the same purpose of simplifying the system basis much in the same way as Row-Echelon forms do for linear systems (The Gr\"obner basis for a linear system is the Row-Echelon form). The Buchberger algorithm performs polynomial division using terms from the old basis in a ranked order. 

\subsubsection{Identifiability in SIR-model}
The relevant differential polynomials for a SIR-model with infected individuals as output measurement are ordered according to rank. $\dot{S}$ and $\dot{I}$ from equation \ref{eq:SIR_model} along with $y = I$ are used to form the polynomials, and $\begin{bmatrix} x_0\\x_1\end{bmatrix} \triangleq \begin{bmatrix} S \\ I \end{bmatrix}$. $A_j^{(k)}$ denotes the $k$th derivative of $A_j$, and the total population $N_{pop}$ is fixed initially to $1$, and the system has no control input.

\textbf{Note:} This example strictly follows the workflow of DAISY, in which internal states and measurements are separately declared. It would be simpler to substitute $x_1$ with $y$ initially when solving by hand.

\begin{align}
    y &- x_1 &\text{leader }x_1\nonumber \\
    \dot{x}_0 &+ \beta x_0x_1  &\text{leader }\dot{x}_0 \label{eq:SIR_Polynomials_first}\\
    \dot{x}_1 &- \beta x_0x_1  &\text{leader }\dot{x}_1 \nonumber
\end{align}
Setting $A_j \triangleq y - x_1, A_i \triangleq \dot{x}_1 - \beta x_0x_1$ to perform division $A_i\div A_j^{(1)} \triangleq R_0$, followed by $R_0 \div A_j^{(0)} \triangleq R_1$. $R_1$ is now reduced with respect to $A_i$, and replaces polynomial 3 in equation \ref{eq:SIR_Polynomials_first}, resulting in the following set of polynomials:
\begin{align}
    \dot{y} &- \beta x_0y + \alpha y &\text{leader } x_0\nonumber\\
    y &- x_1 &\text{leader }x_1 \label{eq:SIR_Polynomials_first_2}\\
    \dot{x}_0 &+ \beta x_0x_1  &\text{leader }\dot{x}_0  \nonumber
\end{align}
Repeatedly applying the pseudodivision algorithm on the polynomials and replacement results in the autoreduced, characteristic set (Solved with DAISY).
\begin{align}
    \Ddot{y}y &- \dot{y}^2 + \dot{y}y^2\beta + y^3\alpha\beta\nonumber\\
    \dot{y} &- x_0y\beta + y\alpha \label{eq:SIR_characteristic}\\
    -x_1 &+ y \nonumber
\end{align}

Polynomial 1 of equation \ref{eq:SIR_characteristic} is the (input)-output relation of the system. The coefficients of this polynomial are extracted and evaluated at symbolic parameter value $p = [1,2]$:
\begin{equation}
    \beta = p_1,\quad \alpha\beta = p_0p_1 \rightarrow \begin{cases}
    \alpha = 1\\ \beta = 2 \end{cases}
\end{equation}
Thus, the parameters have a unique solution and are globally identifiable.
\subsubsection{Identifiability of Basic Models}
\begin{table}[h]
    \centering
    \begin{tabular}{|c|c|c|c|c|c|}
    \hline
    \textbf{Model} & \textbf{Parameters} & \textbf{Inputs} & \textbf{Outputs} & \textbf{Unknown Initial States} & \textbf{Identifiability}\\\hline
        SIR &  $\alpha, \beta$ & None & $I$ & All & Global\\\hline
        SEIR & $\alpha, \gamma$/$\alpha, \beta, \gamma$ & $\beta$/None& $I$ & $(E_0, I_0), I_0$ or $E_0$ & Local\\\hline
        SEIR & $\alpha, \beta, \gamma$ & None& $I$ & None & Global\\\hline
        SEIAR & $\alpha,\beta, \gamma, \mu, p$/$\alpha, \gamma, \mu, p$ & $\beta$/None & $S/E/R$, $I$ &None & Not Identifiable\\\hline
        SEIAR & $\alpha,\beta, \gamma, \mu, p$/$\alpha, \gamma, \mu, p$ & $\beta$/None & I or A & None & Not Identifiable\\\hline
        SEIAR & $\alpha,\beta, \gamma, \mu, p$/$\alpha, \gamma, \mu, p$ & $\beta$/None & $I,A$ & $E_0,I_0$ or $A_0$ & Local \\\hline
        SEIAR & $\alpha,\beta, \gamma, \mu, p$/$\alpha, \gamma, \mu, p$ & $\beta$/None & $I,A$ & None & Global\\\hline
        
    \end{tabular}
    \caption{Identifiability of epidemiological models under different input-output configurations using DAISY}
    \label{tab:Model_Identifiability}
\end{table}
\iffalse
        SIRD & $\alpha, \beta, \nu$ & None & $I$ & None & Not Identifiable\\\hline
        SIRD & $\alpha, \beta, \nu$ & None & $I,D$ & All & Global\\\hline
\fi

The full parameter sets for SIR-variants are identifiable as long as the the number of infected are measured, regardless of initial conditions.

For SEIR-variants the parameter sets also become identifiable when $I$ is measured, but initial conditions are required to globally identify the parameters. This is not a problem in practice, since the second parameter solution has negative, infeasible parameters. 

For SEIAR-variants an additional measurement is required to distinguish the asymptomatic individuals, which makes it possible to identify $p, \mu$ separately from $\beta, \alpha$. The feedback loops in the SEIAR-model also require initial states in order to identify unique parameters globally.

\subsubsection{Identifiability of Other Models}
\begin{figure}[H]
    \centering
    \includegraphics[width=.8\textwidth]{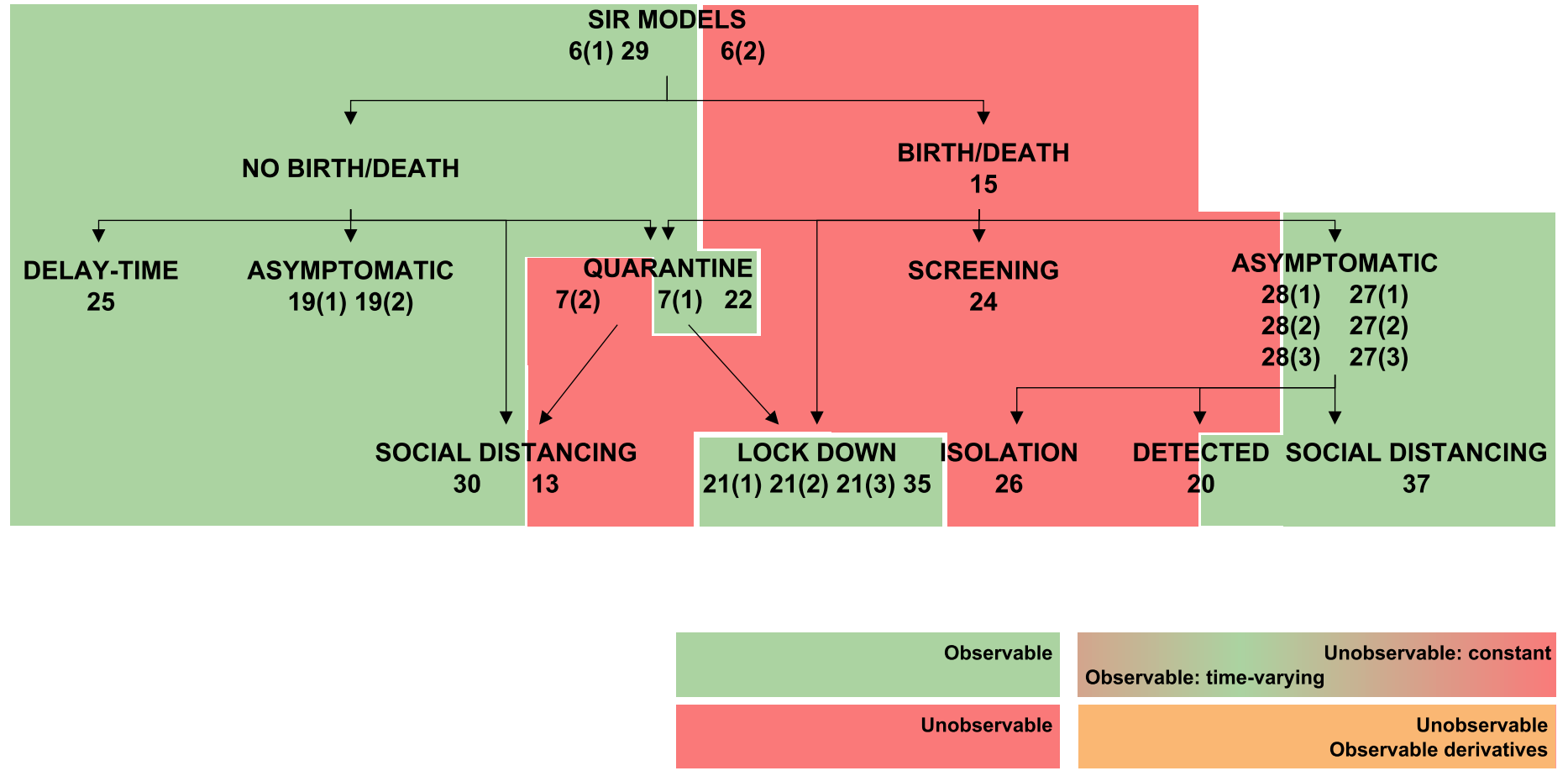}
    \caption{Identifiability of $\beta$ in SIR-models [\cite{massonis2020structural}]}
    \label{fig:SIR_massonis}
\end{figure}

[\cite{massonis2020structural}] investigates identifiability of a large group of compartmental models, which also includes time-varying parameters. Figure \ref{fig:SIR_massonis} shows one of the parameter charts.

The IDs in the charts refer to specific models listed in tables which can be used to quickly find requirements for identifying a chosen model, or used in the model design process to ensure that the model to be used has feasible properties. 

\section{Observability}
While identifiability is important to determine uniqueness of parameters, observability ensures that it is possible to obtain information about the states of a system. Determining the observability (and the dual controllability, which will not be covered) of states is important before considering optimal control and system identification. 
\subsection{Weak Local Observability}
The local observability of a model can be determined using Lie derivatives of the measurement function $h(x)$ and differential equations $f(x)$ of the model. The Lie derivatives are used to form a nonlinear observability matrix for the model, which makes it possible to determine local observability.
\begin{align}
    \dot{x}(t) &=  f(x(t), u(t)), & h(x) &= [y_0, y_1, \dots, y_N]^T
\end{align}
The first Lie derivative is defined as $L_f^0 h(x) = h(x)$, the next are obtained according to equation \ref{eq:Lie_der}.
\begin{equation}
    L_f^{k}h(x) = \frac{\partial}{\partial x}(L_f^{k-1}h(x))f(x)
    \label{eq:Lie_der}
\end{equation}

The observability matrix contains gradient directions of the Lie derivatives and measurements $h_i(x)$ in $h$.

\begin{equation}
    \mathcal{O} = 
    \begin{bmatrix}
    \nabla L_f^0h_0 & \nabla L_f^1h_0 & \dots & \nabla L_f^kh_{0}\\
    \nabla L_f^0h_1 & \dots & \dots& \nabla L_f^kh_{1}\\
    \vdots & \vdots & \vdots & \vdots \\
    \nabla L_f^0h_{n-1} & \nabla L_f^1h_{n-1} & \dots & \nabla L_f^kh_{n-1}
    \end{bmatrix}
\end{equation}
 [\cite{ref:Nonlinear_local_observability}] defines a system to be locally weakly observable  at a given state $x_k$ if $x_k$ is distinguishable from other states $x_i$ in an open neighborhood around $x_k$. This is fulfilled by having linear, independent rows in $\mathcal{O}$.

\subsection{Weak, Local Observability in SIR-model}
Using the system defined in equation. 

\ref{eq:SIR_model} with $h(x) \triangleq I$ results in the following Lie derivatives.
\begin{align}
    L_f^0h(x) &= I\\
    L_f^1h(x) &= \beta \frac{SI}{N_{pop}}
\end{align}
Calculating $L_f^2h(x), L_f^3H(x), \dots$ results in terms containing $S$ and $I$, but not $R$. It is not possible to observe the recovered state from $I$, but it can be determined given initial condition $R(0)$. 

\begin{equation}
    \mathcal{O} = \begin{bmatrix}
    0 & \beta \frac{I}{N_{pop}}\\
    1 & \beta \frac{S}{N_{pop}}
    \end{bmatrix}
    \label{eq:observ_mat_SIR}
\end{equation}

The resulting observability matrix in equation \ref{eq:observ_mat_SIR} remains full rank as long as $I \neq 0$, implying locally weak observability for states where a disease is present in the population.

 \subsection{Structural Observability}
 
 [\cite{ref:algebraic_observability}, definition 3.4] defines a system to be observable if the complete differential-algebraic information of the system is contained in its system equations, making states obtainable from differential equations from the closed \textit{differential field} of the system.
 
 In simpler terms, if it is possible to 'reach' all states $x$ of the system by time differentiating system equations (e.g. Lie derivatives on $h(x)$, equation \ref{eq:Lie_der}) for all points in the state-space, the system is considered to be algebraically observable. 

\subsection{Observability of other Models}
\begin{table}[h]
    \centering
    \begin{tabular}{|c|c|c|c|}
    \hline
    \textbf{Model} & \textbf{Parameters} & \textbf{Outputs}
    & \textbf{Observability}\\\hline
        SIR &  $\alpha, \beta$ & $(I, R)/(I,R,D)$ & Observable\\\hline
        SEIR & $\alpha, \beta, \gamma$ & $(I,R)$ or $(E,R)$ & Observable \\\hline
        SEIA-$R_I$ & $\alpha,\beta, \gamma, \mu, p$ & $(I, R_I)$ or $(E, R_I)$ & Not observable \\\hline
        SEIAR & $\alpha,\beta, \gamma, \mu, p$ & $(I, R)$ or $(E, R)$ & Observable \\\hline
        
    \end{tabular}
    \caption{Observability of epidemiological models using DAISY}
    \label{tab:Model_Observability}
\end{table}

Models where the population transition into 'stable' endpoint compartments (e.g. compartment R) these have to be explicitly measured in order to observe all states. For models with infection feedback-loops (SEIR-variants) it is sufficient to measure from one of the compartment in the closed loop, in addition to the 'stable' comparment. 

For models where the population transition into separate, isolated series of compartments (asymptomatic variants) the observability is in general lost. The same way as with the EI-feedback loop the EA-feedback also require a measurement to distinguish the asymtomatic state from the infectious. In the $SEIAR$-result in table \ref{tab:Model_Observability} this information is avaiable from compartment $R$, which is the end-state for both $I$ and $A$. When only the recovered infectious population is measured ($R_I$) the asymptomatic population is unobservable.

\section{Bayesian Inference}
With increasing availbaility of computation power, bayesian inference has become a very useful process for determining probability distributions of parameters for complex systems where analytical solutions are difficult to obtain. Bayesian inference is a process where a prior belief is updated as information becomes available, resulting in a posterior distribution.

The major advantage with Bayesian inference is that it makes it possible to obtain (probability-distributed) parameter solutions without analytically solving complex equations. The introduced SIR, SEIR and SEIAR Chain-Binomial models are examples of such models. The posterior probabilities of the infection transmission alone can become very complex after few time iterations, even for small population groups [\cite{ref:BinChain}].

\subsection{Bayes Rule}
Following a similar notation to [\cite{ref:SMC-methods}] the posterior distribution can be expressed in terms of Bayes rule:
\begin{equation}
    p(x_{0:k} | y_{1:k}, \theta) = \frac{p(y_{1:k} | x_{0:k}, \theta)p(x_{0:k} | \theta)}{p(y_{1:k} | \theta)} \propto p(y_{1:k} | x_{0:k}, \theta)p(x_{0:k} | \theta)
    \label{eq:BayeInf_Prop}
\end{equation}

Where $x_{0:k}$ is the state trajectory, $y_{1:k}$ are observations and $\theta$ is the model parameters. For system identification the high probability regions of the posterior can be obtained by evaluating the trajectories of the system that are most probable.

\subsection{Sequential Monte Carlo}
Sequential Monte Carlo (SMC, also called particle filters) is used to explore the outcome space for the most probable trajectories according to a given prior distribution and likelihood function. Particles are used in differential equations to evaluate trajectory steps, and the state of the particles are used to assign individual weights. The weight of a particle describes how 'relevant' its trajectory is to given measurements, but each step also gives an indication of how well the assigned proposal parameters $\theta_{prop}$ performs. 

It is not desired to evaluate the trajectory of 'unrelevant' particles. When the weight of particles become small compared to others they contribute less to shaping the posterior density. The following sections address this, in addition to the case where particles uniformly become too 'relevant'.

\subsubsection{Importance Sampling}
The posterior density $p(\theta | Y)$ is only known to proportionality (equation \ref{eq:BayeInf_Prop}), but it is possible to approximate the statistical moments of the distribution with a sufficient number of samples. The statistical moments of the posterior can be extracted by integrating the product of posterior $p(x_{0:k}|y_{1:k}, \theta)$ and a function of interest $f_t$.

\begin{equation}
    \mathbb{E}[f_k(x_{0:k})] = \int f_t(x_{0:k}| \theta)p(x_{0:k}|y_{1:k}, \theta)d x_{0:k}
    \label{eq:Info-Integral_impoartance}
\end{equation}  One useful example of $f_t$ is the conditional mean $f_t(x_{0:t}) = x_{0:t}$ which makes it possible to obtain a posterior mean using the integral. Instead of using sampling averages directly from $p(x_k|y_1, \dots, y_k, \theta)$, an importance distribution $\pi(x_{0:k}, y_{1:k})$ is introduced. Equation \ref{eq:Info-Integral_impoartance} can be reformulated in terms of importance weights.
\begin{align}
    \mathbb{E}(f_k) &= \frac{\int f_t(x_{0:k} | \theta)w_k(x_{0:k} | \theta)\pi(x_{0:k}|y_{1:k}, \theta)dx_{0:k}}{\int w(x_{0:k} | \theta)\pi(x_{0:k}|y_{1:k}, \theta)dx_{0:k}}\\
    w(x_{0:k} | \theta) &= \frac{p(x_{0:k} | \theta)}{\pi(x_{0:k}|y_{1:k}, \theta)}
\end{align}

\subsubsection{Sequential Importance Sampling}
\label{ch:SIS_SMC}
Importance sampling is a variance reduction technique where samples are selected with respect to a different probability distribution. The Sequential Importance Sampling (SIS) approach is one of the simplest, and can be summarized in three steps. 
\begin{itemize}
    \item Initialize with $x_0$, parameter proposal $\theta$ and measurements $y$.
    \item For each iteration: use system dynamics $f(\hat{x}_{k-1}, \theta)$ to obtain state estimates $\hat{x}_k$. Use the likelihood of $y_k$ ($p(\hat{x}_k | y_k)$) to assign weights to the estimates.
    \item Select new samples with the normalized weights as probability mass function.
\end{itemize}

The weight calculation can be written as a recursive step:
\begin{equation}
    w_k \propto w_{k-1} \frac{p(y_k | x_k)p(x_k | x_{k-1})}{\pi(x_k | x_{0:k-1}, y_{1:k})}
\end{equation}

This is a simple approach to reducing variance and exploring the probable regions of the state space. However, this approach may lead to \textit{degeneracy} where all weight are assigned to few/a single particle. When degeneracy occurs most of the particles explore the same region of the state space, which is the opposite of what the particles are intended to do.
\subsubsection{Sequential Importance Sampling and Resampling}
Sequential Importance sampling and Resampling (SIR) solves the problem with degeneracy in SIS-resampling by introducing a measure for the effective number of particles:
\begin{equation}
    N_{eff} = \frac{1}{\sum_{i=1}^{N_p}(w_k^{(i)})^2}
\end{equation}
Where $N_p$ is the total number of particles. If $N_{eff}$ falls below a given threshold, the particles are resampled according to their current weights, followed by all weights being set uniformly to $\frac{1}{N_p}$.

\subsection{Markov-Chain Monte Carlo}
Markov-Chain Monte Carol (MCMC) is a widely used approach for estimating model parameters for systems where true states are not directly observable, but obtainable from probability distributed variables. The method applies Sequential Monte Carlo simulations to Markov models in order to determine the posterior probability distribution of the parameters to be identified. 

\subsubsection{Metropolis Algorithm}
\label{Ch:Metropolis}
When a trajectory with weights has been computed (section \ref{ch:SIS_SMC}) the weights can be used to evaluate if the proposed parameters $\theta_i$ were better or worse than the previous proposal $\theta_{i-1}$. The Metropolis algorithm is one of many solutions for determining the sequence of parameters to evaluate, and is applied and evaluated over multiple trajectories.

\begin{algorithm}[H]
\SetAlgoLined
\KwData{$\theta_i = \theta_0$ where $p(\theta_0 | y) > 0$, $N_\theta$}
\For{i in $N_\theta$}{
    Evaluate trajectory for $\theta_i$\\
    Retrieve the average weight $\frac{\sum_{k=1}^{N_t} p(\hat{x}_k | y_k)}{N_t} \approx p(\theta_i | y)$ \\
    Calculate $\alpha = \min(1, \frac{p(\theta_i | y)}{p(\theta_{i-1} | y)})$\\
    \If{Uniform$(0,1) > \alpha$ and $\theta_i \neq \theta_0$}{ draw $\theta_{i+1}$ from the symmetric distribution $J(\theta_{i+1} | \theta_{i-1})$}
    \Else {draw $\theta_{i+1}$ from the symmetric distribution $J(\theta_{i+1}|\theta_i)$}
 }
 \caption{SMC with Metropolis Algorithm}
 \label{alg:SMC_Metropolis}
\end{algorithm}
The Metropolis algorithm works only for symmetric proposal distributions $J$.
\subsubsection{Metropolis-Hastings}
To balance out the likelihood ratio $\alpha$ (algorithm \ref{alg:SMC_Metropolis}) with respect to the asymmetry of $J$, the likelihoods are corrected using proposal distribution $J$ (equation \ref{eq:hastings})

\begin{equation}
    \alpha = \frac{\frac{p(\theta_i | y)}{J(\theta_i | \theta_{i-1})}}{\frac{p(\theta_{i-1}|y)}{J(\theta_{i-1} | \theta_i)}}
    \label{eq:hastings}
\end{equation}

The use of asymmetric proposal distributions can improve parameter proposals and make the algorithm find high parameter likelihoods faster. Given that the parameter proposals sufficiently explore the parameter space, the empirical distribution of saved parameters $\theta_0, \dots, \theta_N$ will converge to the true parameter distribution.

\subsection{Implementations: PyMC3}
SMC and MCMC-algorithms can require a lot of evaluations in order to produce good posterior densities. This makes it important to choose a language which supports fast distribution sampling. Along with simpler, custom implementations of MCMC and SMC, libraries in both Python and C++ have been considered.

PyMC3 [\cite{ref:pyMC3}] is a library in Python for bayesian statistical modeling in general, and supports the use of SMC and MCMC-samplers. The samplers have been able to evaluate model dynamics effectively through the use of Theano, a library which uses an optimizing compiler to translate algebraic expressions to code efficiently, with the addition of GPU-support. 

The development of Theano was officially ended in 2018 but forked and to some extent maintained by the pyMC-team. A substantial effort was made in order to setup and use the pyMC3 interface for this project, which eventually succeeded for one specific version of Python (3.8) on Ubuntu, but no solution was found compatible for Windows.

The pyMC3 application interface provides good solutions for continous ODEs and Markov chains with state-invariant transition-distributions, but did not support binomial sampling for ODEs directly. An attempt was made at solving the issue by constructing a black-box Theano model, but this breaks the intended abstraction of the library, and did not result in good performance either.

\subsection{A Custom SMC/MCMC-Library}
In order to remove the abstraction and get an interface able to utilize hardware better, a SMC-library called Sequential Monte Carlo Template Class [\cite{SMCTC}] was redesigned. The library takes advantage of the templating function in C++ in order to generate code for a sampler fitted to a custom particle class, with custom system dynamics and probability densities.

\begin{figure}[H]
    \centering
    \includegraphics[width=.8\textwidth]{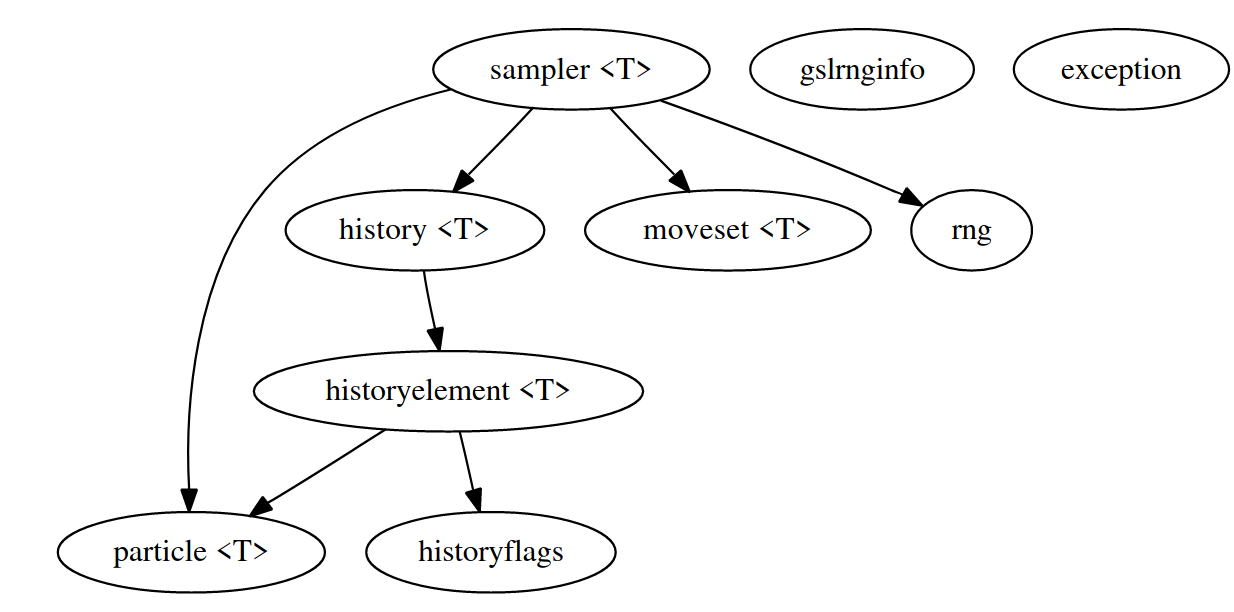}
    \caption{SMCTC Collaboration Chart[\cite{SMCTC}]}
    \label{fig:SMCTC_Collab}
\end{figure}
The library depends on the GSL-library[\cite{ref:GSL_lib}] for random number generation, which samples substantially faster than Python's NumPy and SciPy implementations. 

\subsubsection{Overview}
The custom implementation makes use of some features from the sampler, rng and particle classes, but is modularized differently in order to clearly separate between data storage and routines. 

\begin{figure}[H]
    \centering
    \includegraphics[width=.6\textwidth]{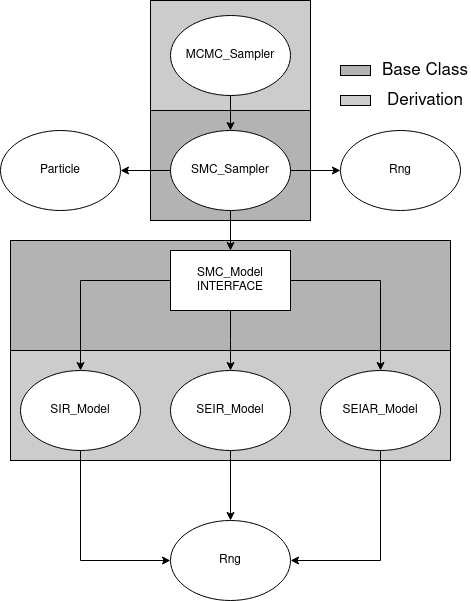}
    \caption{Collaboration chart for custom MCMC-library}
    \label{fig:CollabCustomMCMC}
\end{figure}

\subsubsection{Interface}
An interface class \textit{SMC\_Model.hh} is implemented to limit the interaction between the SMC/MCMC-sampler and the custom model implementation. The following virtual functions listed in the class needs to be implemented by the user:

\lstinline[language=C]{void Init(double* &, long &)}

Allocate space needed and initialize state.

\lstinline[language=C]{void Step(const long &lTime, double* X, const double *param)} 

Move the particle state to the next timestep.

\lstinline[language=C]{void ProposalSample(const double* oldParam, double* result)} 

Draw a new parameter proposal from a custom distribution.

\lstinline[language=C]{double LogLikelihood(const double *X, const long &lTime)} 

Evaluate likelihood for the current timestep using current estimate $X$ and the observation data.

\lstinline[language=C]{void Reset(double* X)} 

Reset state values in order to prepare for a new SMC-trajectory evaluation.

\subsubsection{SMC Sampler}
The SMC-sampler sequentially pass particle states and weights to the model, and afterwards perform resampling. Its workflow can be briefly summarized in a few declarations:

\lstinline[language=C]{void InitializeParticles()}

Initializes the state and weight of all particles.

\lstinline[language=C]{void MoveParticles()}

Moves all particles to the next timestep.

\lstinline[language=C]{double GetESS()}

Calculate the current Effective Sample Size.

\lstinline[language=C]{void Resample(ResampleType lMode)}

Performs resampling using the current particle weights, with support for multiple resampling strategies.

\lstinline[language=C]{void Normalize_Accumulate_Weights()}

Accumulates the weights throughout the SMC-run, and normalizes the weights to sensible values for importance and resampling.

\lstinline[language=C]{void ResetParticles()}

Set state and weights of each particle back to initial.

\subsubsection{MCMC Sampler}

The MCMC-sampler controls the SMC-iterations and keeps track of the used parameters and accumulated weights. It can either be constructed independently, or built on top of an SMC sampler using a copy constructor.

\lstinline[language=C]{void IterateMCMC()}

Runs SMC sampler iterations until it has reached end of trajectory. Afterwards \lstinline[language=C]{Metropolis()} is called to propose a new parameter, followed by a particle reset. 

\lstinline[language=C]{int Metropolis(void)}

Runs Metropolis algorithm (section \ref{Ch:Metropolis}) with current and previous weights and parameters.

\textbf{Note:} Only symmetric proposal distributions have been used in this project. Metropolis-Hastings requires a small extension to the interface and MCMC\_Sampler classes, but can easily be added when needed. 

\subsubsection{Comparison of Performance}
The stochastic trajectory simulations in Python were timed to use an average of $\approx 0.135s$ per SIR trajectory when evaluating beta-binomial distributions with the Scipy library. The custom MCMC-implementation was measured to evaluate 100 particle trajectories with resampling and weight updates in $\approx 5 \pm 1 ms$.

\subsection{Simulation Setup}
Given a list of measured infections and an initial state, there are many parameter sets $\theta$ for the stochastic SIR, SEIR and SEIAR models which are able to produce the same results.  This section presents a method to sample from the epidemiological Chain-Binomial models, which will be applied on measurements generated by the following deterministic models.

\begin{figure}[H]
    \centering
    \includegraphics[width=.8\textwidth]{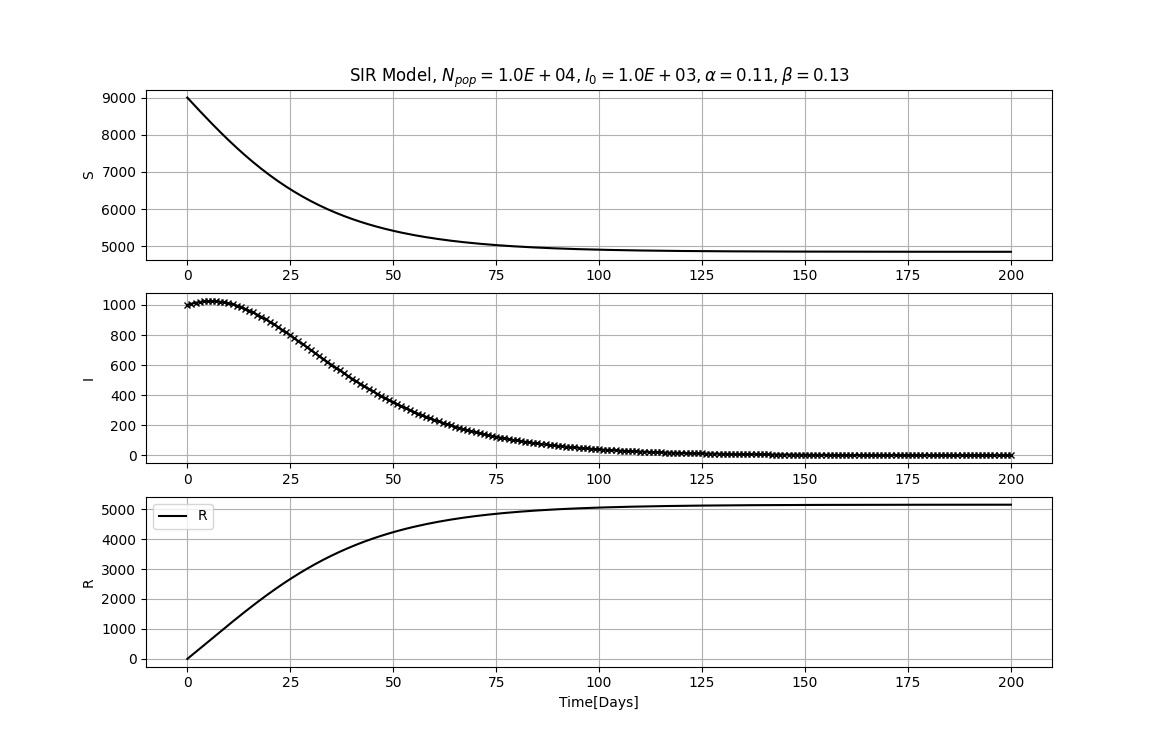}
    \caption{Deterministic SIR-trajectory for measurements}    
    \label{fig:Problem_Trajectory_SIR}
\end{figure}

\begin{figure}[H]
    \centering
    \includegraphics[width=.8\textwidth]{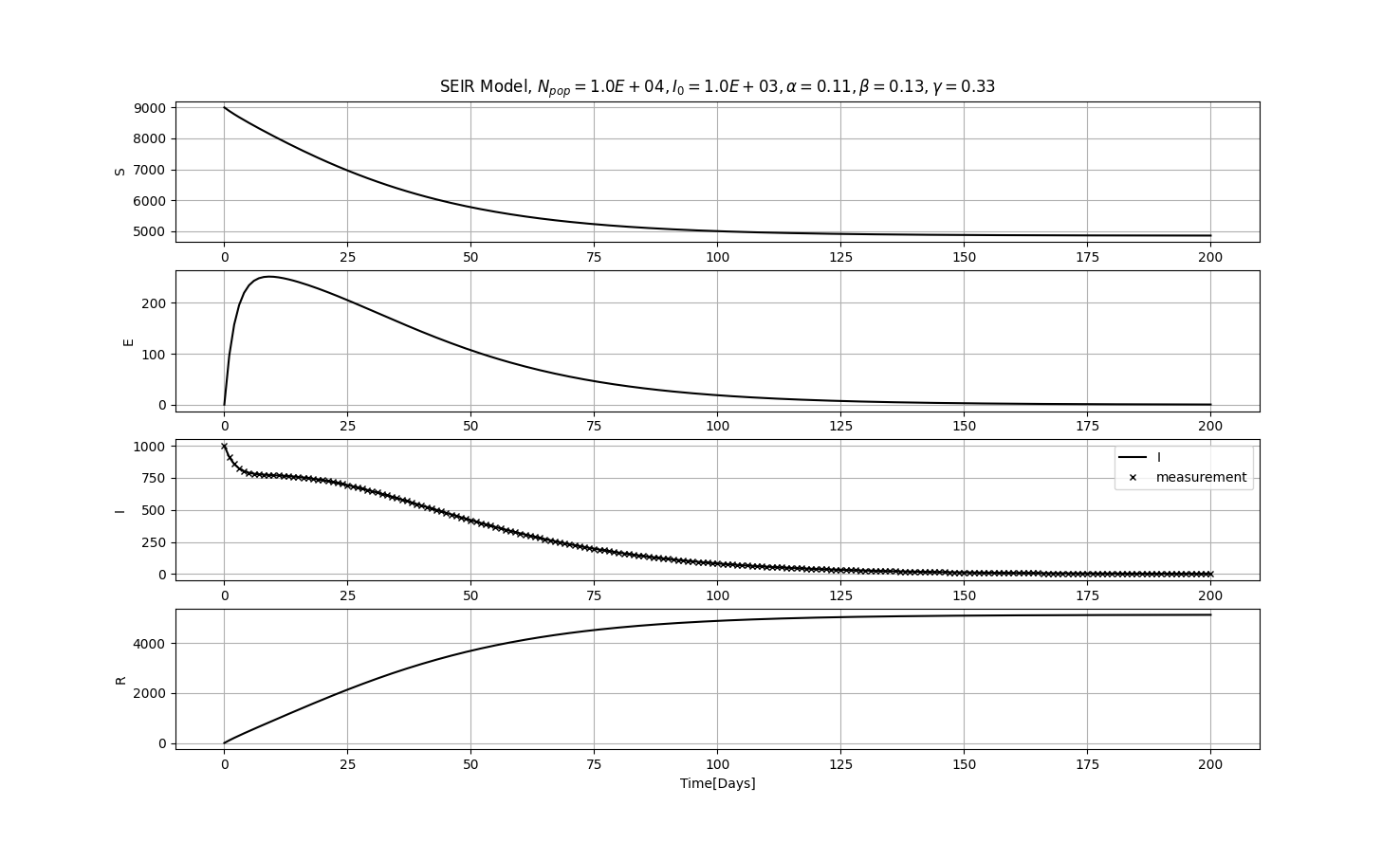}
    \caption{Deterministic SEIR-trajectory for measurements}    
    \label{fig:Problem_Trajectory_SEIR}
\end{figure}

\begin{figure}[H]
    \centering
    \includegraphics[width=.8\textwidth]{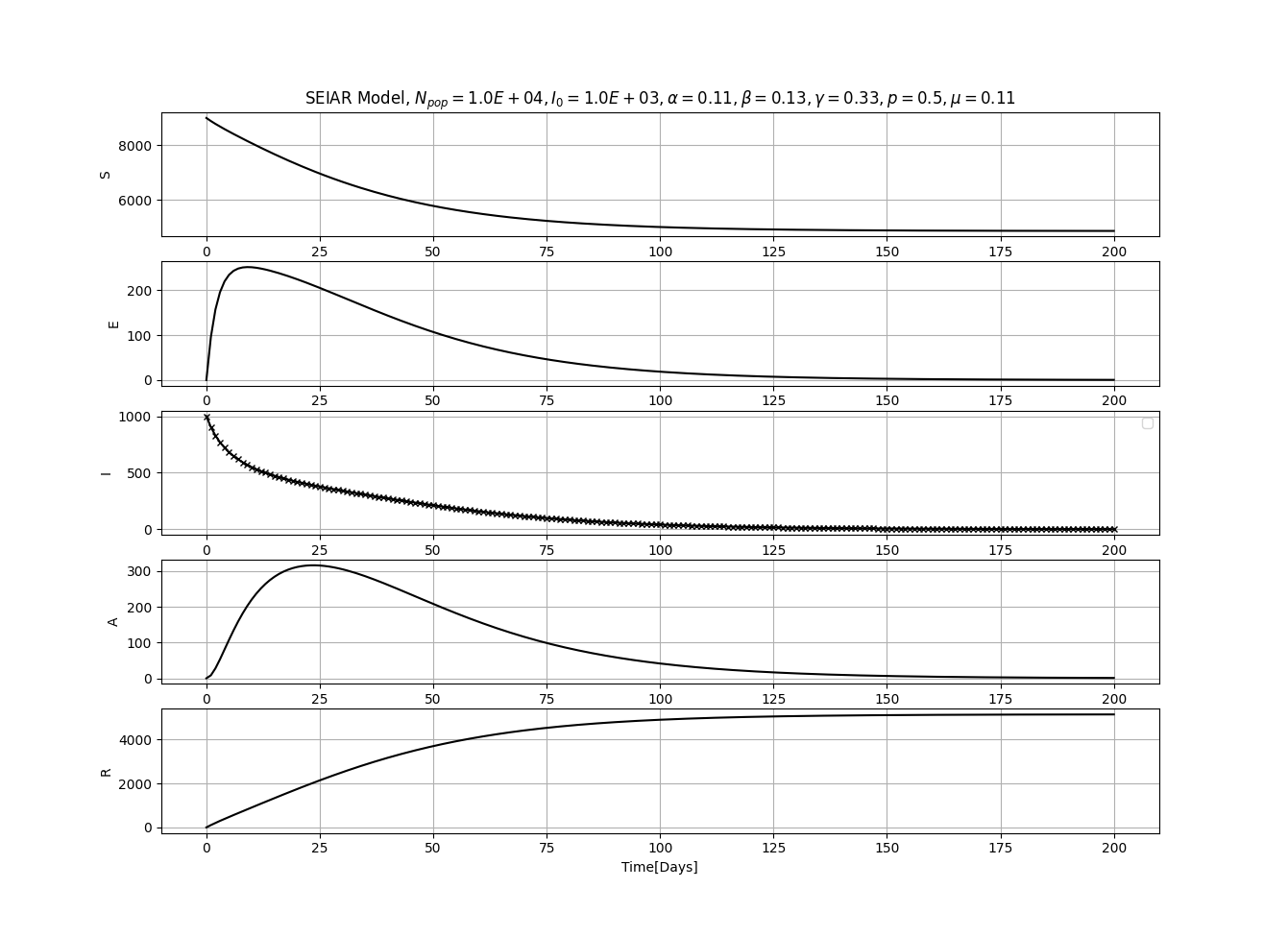}
    \caption{Deterministic SEIAR-trajectory for measurements}    
    \label{fig:Problem_Trajectory_SEIAR}
\end{figure}

The identification algorithm presented in this section will be given the initial conditions $x_0$, total population $N_{pop}$ and step size $\Delta t$, in order to evaluate which parameter set $\hat{\theta}$ that is most likely to reproduce the given measurements using a stochastic model. Initial parameter proposal $\theta_{prop, 0}$ will be set to $2\theta$ (Where $\theta$ is the true, deterministic parameter set, and with the exception of parameter $p$, which is set to $1.5p_{true}$). 

\subsection{Simulation Results}

Running the implemented MCMC algorithm with the stochastic SIR-model, 100 particles, 50000 iterations, $p(x | \theta, y) = \mathscr{N}(x;\sigma=100)$ and proposal parameter $\theta_{prop, k} \sim \mathscr{N}(\theta_{k-1}, \begin{bmatrix} 0.1\alpha & 0.1\beta \end{bmatrix} \cdot I) \in [0,\infty)$ results in the following parameter distributions:

\begin{figure}[H]
    \centering
    \includegraphics[width=.8\textwidth]{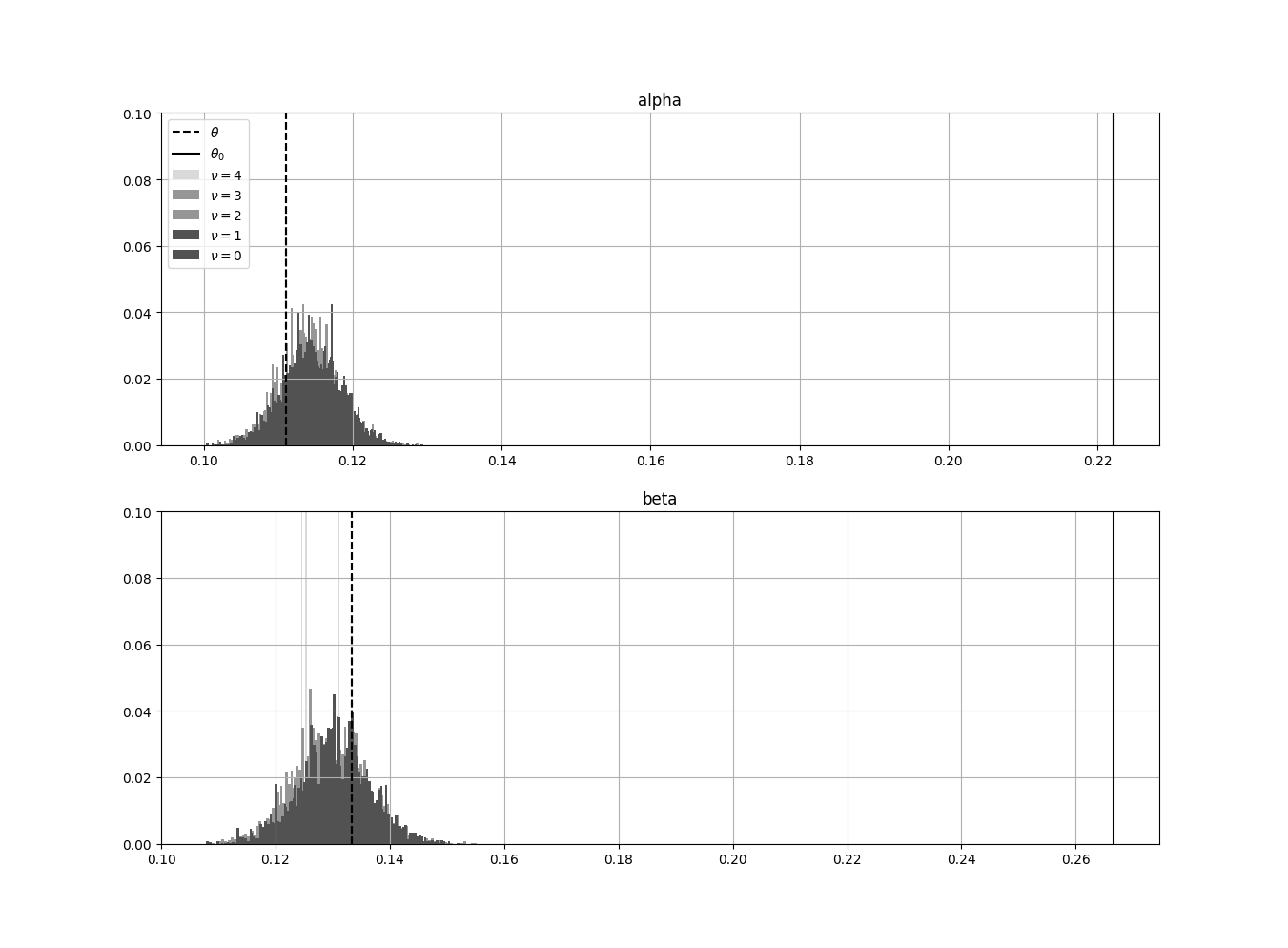}
    \caption{Parameter distributions for Stochastic SIR-Model using implemented C++ MCMC-library.}
    \label{fig:SIR_MCMC}
\end{figure}
\iffalse
Running the implemented MCMC algorithm with the stochastic SIR-model, 100 particles, 50000 iterations, $p(x | \theta, y) = \mathscr{N}(x;\sigma=100)$ and proposal parameter $\theta_{prop, k} \sim \mathscr{N}(\theta_{k-1}, \begin{bmatrix} 0.1\alpha & 0.1\beta \end{bmatrix} \cdot I) \in [0,\infty)$ in an initial outbreak with $I_0=10$ and $\theta_{prop, 0} = \theta$ resulted in the following parameter distributions:

\begin{figure}[H]
    \centering
    \includegraphics[width=.8\textwidth]{Images/SIR_SmallSize_MCMC.png}
    \caption{Parameter distributions for Stochastic SIR-Model in an initial outbreak using implemented C++ MCMC-library.}
    \label{fig:SIR_MCMC}
\end{figure}
\fi
Running the implemented MCMC algorithm with the stochastic SEIR-model, 100 particles, 50000 iterations and $p(x | \theta, y) = \mathscr{N}(x;\sigma=100)$ resulted in the following parameter distributions:

\begin{figure}[H]
    \centering
    \includegraphics[width=.8\textwidth]{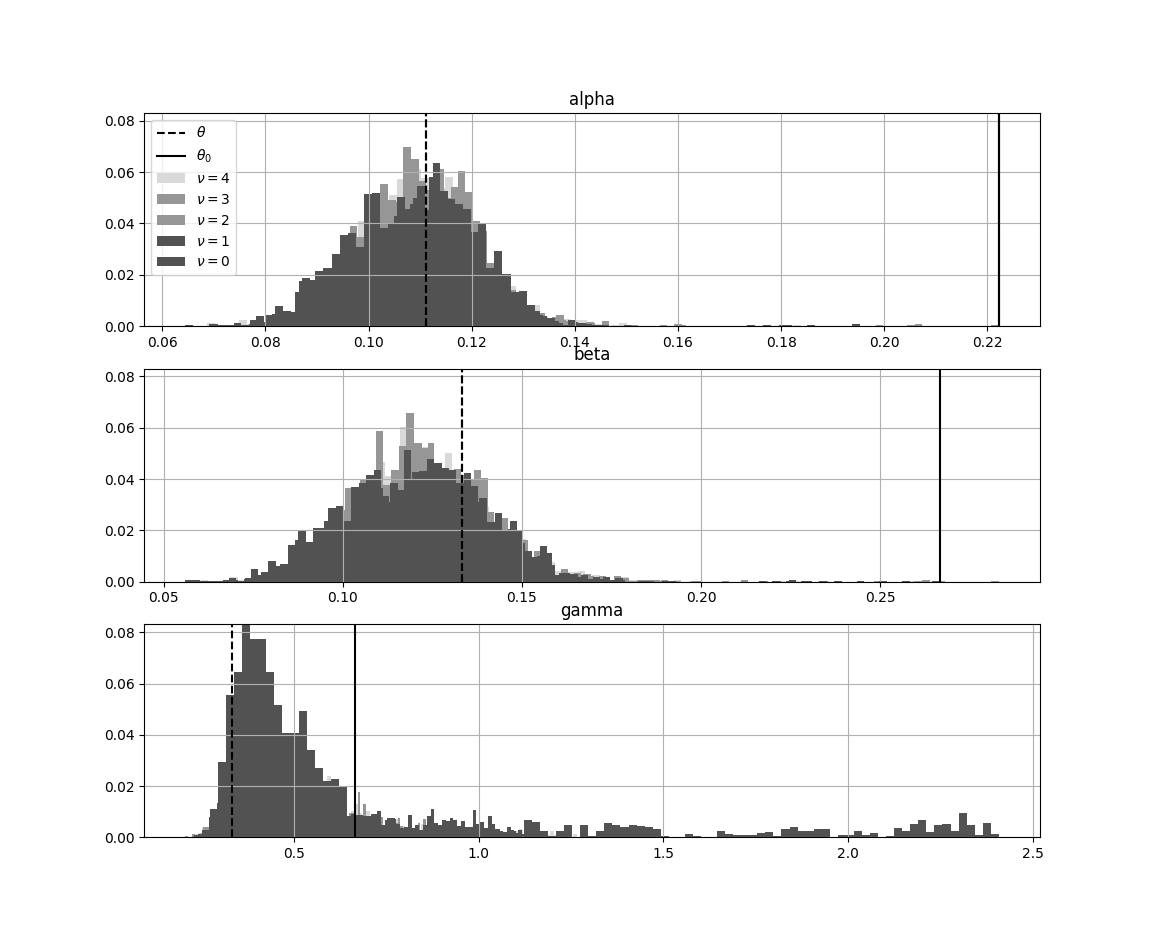}
    \caption{Parameter distributions for Stochastic SEIR-Model using implemented C++ MCMC-library.}
    \label{fig:SEIR_MCMC}
\end{figure}

Running the implemented MCMC algorithm with the stochastic SEIAR-model, 100 particles, 50000 iterations and $p(x | \theta, y) = \mathscr{N}(x;\sigma=100)$ resulted in the following parameter distributions:

\begin{figure}[H]
    \centering
    \includegraphics[width=.8\textwidth]{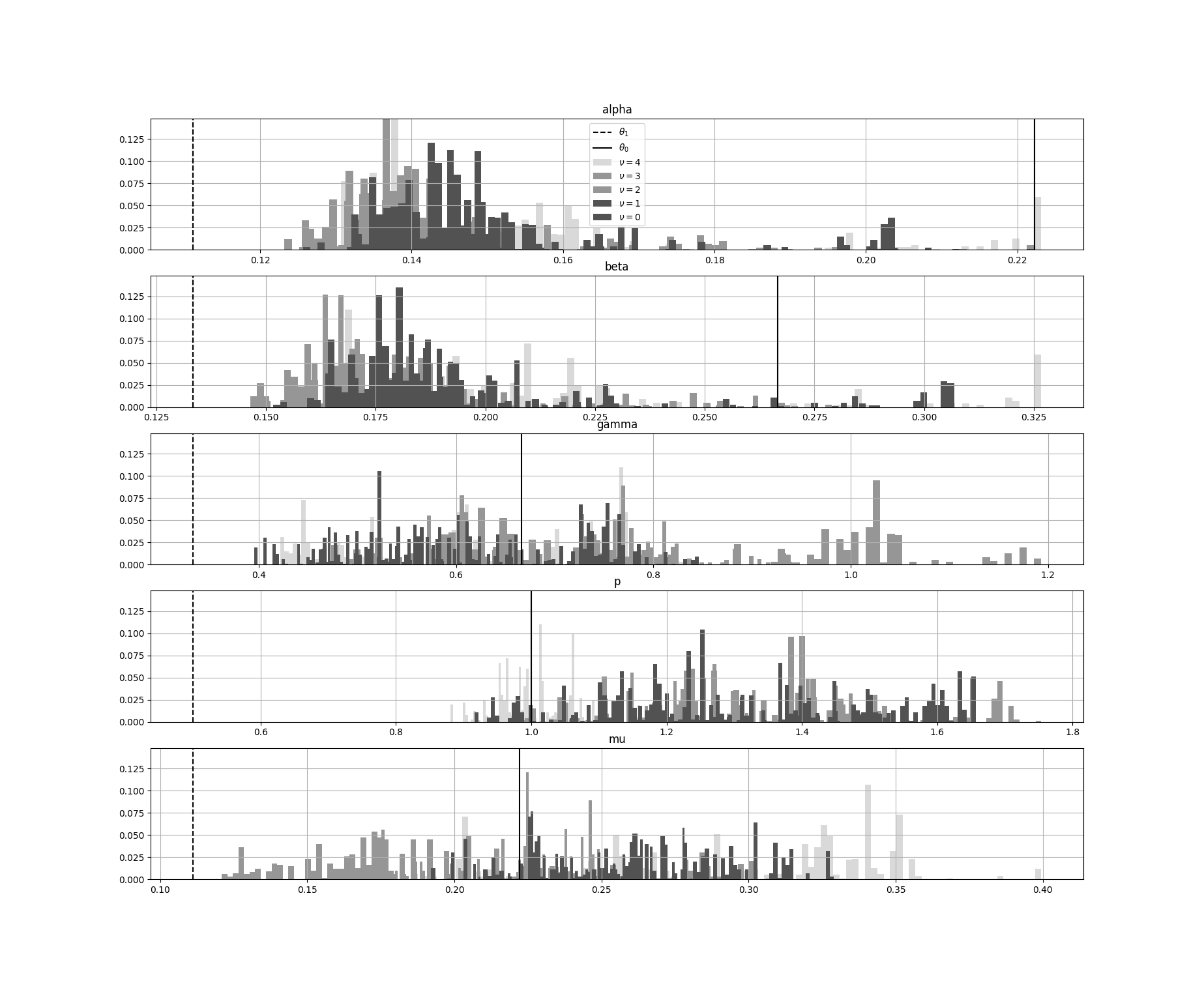}
    \caption{Parameter distribution for Stochastic SEIAR-Model using implemented C++ MCMC-library.}
    \label{fig:SEIAR_MCMC}
\end{figure}

\subsection{Remarks on the Simulations}

There are many possibilities for parameter tuning within SMC/MCMC, where some give good convergence to a true solution, and others lead to no convergence at all. In this project, the algorithm has been tuned from intuitions, which may be impossible for more complex models.  

$p(x_{0:k}|\theta, y_{1:k})$ was tuned to sufficiently distinguish a bad infection count prediction from a good one, in the sense that a 'good' stochastic model has the tendency to estimate within one standard deviation $\sigma$ of $y_{1:k}$. Reducing $\sigma$ could speed up convergence locally (which was not prioritized, given the 50000 iterations used) at the cost slowing down convergence when $\theta_0$ is far from $\theta$. 

Different magnitudes was used for the standard deviation of $\theta_{prop, k}$. Using a large deviation resulted in an increase in Metropolis rejections, since the parameters proposed gave radical trajectories. Using a very small deviation did not ensure a consistent convergence either. Because of the randomness from the stochastic models, an accepted small parameter step do not neccesarily have to be a step in the right direction.

\subsection{Discussion}
The implemented MCMC-library managed to identify a reasonable estimate for the parameter for the  stochastic SIR and SEIR models. The parameter distributions in figures \ref{fig:SIR_MCMC}, \ref{fig:SEIR_MCMC} are mostly normally distributed with an offset from the true deterministic parameters.

The stochastic model require a different parameterization than the deterministic for this population size. This is relatable to the simulated trajectories in section \ref{ch:EpidModels} which show that parameterization at this population size makes the stochastic models behave slightly different than the deterministic ones.  

An increase in $\nu$ did not have a significant impact on the parameter distributions. There are many dependencies in parameterization here, which makes it difficult to determine if this is a valid result. The population size is moderately large ($N_{pop} = 10000)$ and homogenous, which according to results from section \ref{ch:EpidModels} indicates that $\nu$ should have a small impact on the trajectory. This is consistent with the current result. On the other hand, the moderately large standard deviation $\sigma=100$ could have slowed down the convergence towards the true estimate, but 50000 iterations should counteract the impact of this.

\section{Future Work}
\subsection{Networked Mean Field Approximations}
As mentioned in the introduction it is desired to associate the deterministic to the stochastic nature of true epidemics. While Chain-Binomials are able to capture initial outbreaks, they do not use the spatial information required to do accurate predictions.

Instead of modelling initial outbreaks with Chain-Binomials [\cite{Bussell405746}] use deterministic, networked  mean-field approximations ([\cite{nowzari2015analysis}]) to approximate a network with multiple compartmental models. 

\begin{figure}[H]
    \centering
    \includegraphics[width=.6\textwidth]{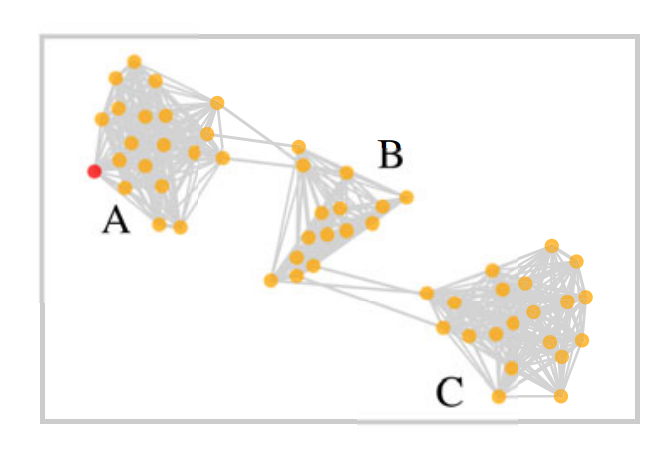}
    \caption{Network subject to mean-field approximations [\cite{Bussell405746}]}
    \label{fig:busselNet}
\end{figure}

Finding good approximations for network clusters, possibly with confidence intervals, would be a step in the right direction, and may be necessary in order to apply optimal control on smaller populations. The results from [\cite{Bussell405746}] show that optimal control with three deterministic approximation for the given network outperforms the alternative choices.

If the custom MCMC-implementation could be improved to be computationally efficient for network models, it could further be possible to use spatial infection measurements to infer posterior parameter distributions for deterministic SIR/SEIR/SEIAR models. 

Local solutions for the deterministic models have already been studied, and are attainable through the use of multiple shooting methods and Interior-Point Optimization (IPOPT) [\cite{Wächter2006}]. This sort of optimization could possibly extended to account for multiple compartmental models, resulting in a robust strategy which integrates spatial information, while still being able to retain anayltical properties in the model.

\subsection{Practical Identifiability and Dynamic Observability}
While structural observability and idenitifiability describes the uniqueness of states and parameters, it does not necessarily hold when the models are subject to noise. In this case it may be a need for a continous quantification of identifiability and observability, which is addressed respectively in [\cite{ref:practical_identificaiton}], [\cite{ref:dyn_obs}].

\subsection{MCMC Parameterization and Correlation}
There are many aspects to be explored within parameterization of the SMC/MCMC-algorithms. Effective sample size $N_{eff}$, the number of SMC-particles and the previously mentioned distributions give many options, and this only covers the methods of standard SIR-resampling and parameter choices with Metropolis. There are many SMC and MCMC-methods to explore, some of them are mentionable.

\begin{itemize}
    \item Hamiltonian MCMC - Models the parameters as a particle, which conserves potential/kinetic energy as it travels through the parameter space, where the energy distribution is determined by the measurement likelihood. 
    \item Gibbs Sampling - A frequently used special case of Metropolis Hastings.
    \item In addition to standard resampling, the SMCTC-library implemented Residual, Stratified and Systematic resampling strategies. 
\end{itemize}

Using old parameters to propose new ones naturally leads to correlated decisions in the MCMC-sampler. Due to the sample size, this was negligible for the project simulations, but its impact should be studied for future projects. 

\subsection{Optimal Control on NTNUs COVID-19 Model}
The model implemented by the NTNU COVID-19 Taskforce use the following model/control scheme:
\begin{figure}[H]
    \centering
\includegraphics[width=.8\textwidth]{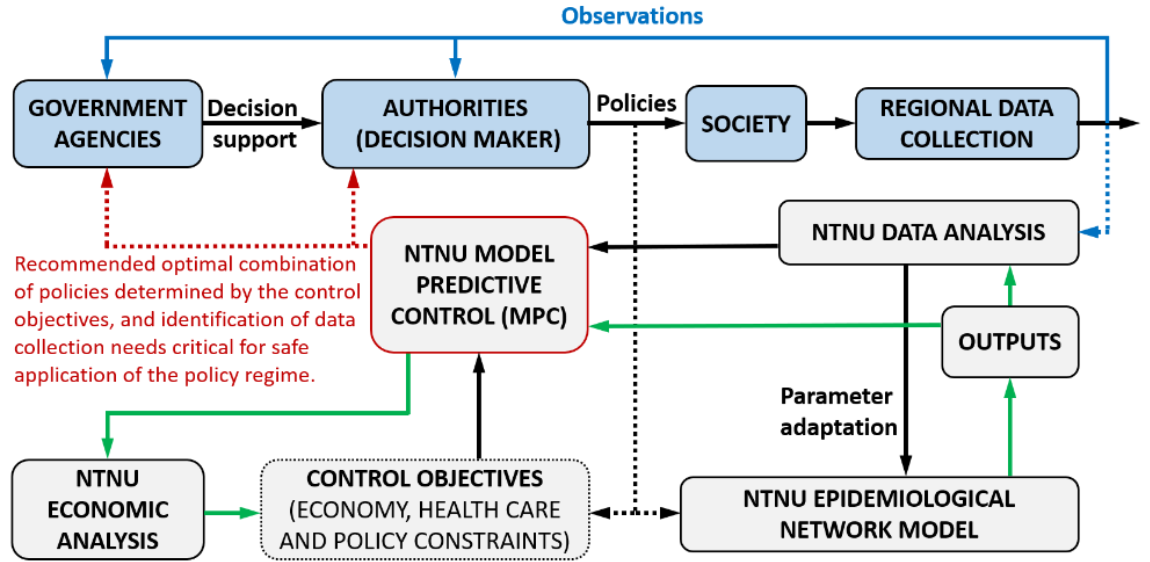}
    \caption{Block Diagram for the NTNU COVID-19 Modeling and Control scheme [\cite{ref:NTNU_COVID}]}
    \label{fig:NTNU_control_scheme}
\end{figure}

The current Model Predictive Control (MPC) use the deterministic SEIR model as approximation for the underlying network model. However, if the network consists of many clusters (figure \ref{fig:busselNet}) it would be better approximated with multiple SEIR-models. Using SMC and MCMC to find parameter distributions for multiple models can lead to a more accurate and robust control strategy.
%\import{./Sections/}{Identification_Methods}
\newcommand{\eqdef}{=\mathrel{\mathop:}}

% Example section added directly into the main-file
\section{Conclusion}

A variety of probability distributions have been studied in this project, providing ways of representing overdispersion in Chain-Binomial models. Onwards,  the overdispersed distributions has been used to illustrate how trajectories change in epidemiological models, which resulted in a moderately weak impact given a total population size of $N_{pop} > 10000$. 

Trajectory simulations show that Chain-Binomials are important for modeling non-homogenous small populations, while it is possible to approximate them with deterministic models for homogenous, large populations. These simulations do show indicative trends, but a larger sample size may be needed to approximate true confidence intervals.

The structural identifiability of the SIR, SEIR and SEIAR models was found in different configurations using DAISY. While parameters in SIR and SEIR are identifiable, additional measures are needed in order to retrieve parameters in models with asymptomatic compartments. 

The structural observability of the SIR, SEIR and SEIAR models was assessed in different configurations, revealing that the asymptomatic compartment is structurally unobservable, while the SIR and SEIR models are structurally observable.

The Sequential Monte Carlo and Markov Chain Monte Carlo methods were studied to find ways of estimating parameter distributions for Chain Binomial models. This resulted in a computationally efficient implementation which successfully converges towards true parameters, given a good parameterization. The exception was the unobservable SEIAR model, which did not converge towards the true parameter values. 

Overall, the project has been a good introduction to clarify what kind of information that is retrievable from the simplest epidemiological models, and it has presented ways of modelling unavailable information in terms of uncertainty,  and it has provided an identification algorithm that can be applied to more complex models.

% Printing bibliography
\newpage
\printbibliography[heading = bibintoc, title = Bibliography]    % 'bibintoc' inserts our bibliography into the table of contents

% Inserting appendix with separate settings
\addappendix
\subsection{Buchberger Algorithm}

Denoting $LM(p)$ as the leader mononial and $LT(p)$ as the leader term of polynomial $p$ ,and $LCM(p,q)$ the least common multiple of polynomials $p, q$, S-polynomials (Subtraction) can be defined:
\begin{equation}
    S(p,q) = \frac{LCM(LM(p), LM(q))}{LT(p)}\cdot p - \frac{LCM(LM(p), LM(q))}{LT(q)}\cdot q
\end{equation}
\begin{algorithm}[H]
\SetAlgoLined
\KwData{Input basis $f = \{f_1, \dots, f_N\}$}
g:= f
\While{g $\neq$ h}{
    h := g
    \For{all pairs $(p,q), p\neq q, p,q\in g$}
    {
    $S$:=$S(p,q)$\\
    $r$:=remainder(S, g)\\
    \If {$r \neq 0$}
        {
        add $r$ to basis $g$
        }
    }
 }
 \caption{Buchberger Algorithm}
 \label{alg:Buchberger}
\end{algorithm}

\section{Code Overview}
Working examples for the MCMC are frozen on the "Inspera" branch of a git repository:

\lstinline[language=bash]{git clone https://github.com/jonasbhjulstad/MCMC.git}

The project is structured with the following top-level folders:
\begin{figure}[H]
    \centering
    \includegraphics[width=.5\textwidth]{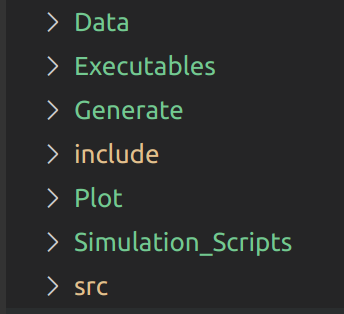}
    \caption{Top-level folders of C++ project}
    \label{fig:folderstruct}
\end{figure}
\textbf{Note:} The file storage paths must be updated in order to direct the parameter/weight output to correct filepaths. 

\begin{itemize}
    \item Data - Stores parameters, weights and generated trajectories
    \item Executables - Contains the compiled binaries. (run\_SXXX, gen\_XXX)
    \item Generate - Contains source files for generating trajectories
    \item Plot - Contains .py-files for trajectory and parameter plots
    \item Simulation\_Scripts - Contains One main parameter configuration file, and separate execution loops for the models.
    \item src - Contains source code files
\end{itemize}
\end{document}